# Direct Simulation Monte Carlo Analysis on Thrust Vectoring of a Supersonic Micro Nozzle using Bypass Mass Injection

Maruf Md. Ikram, Abu Taqui Md. Tahsin, and A. B. M. Toufique Hasan*

Department of Mechanical Engineering, Bangladesh University of Engineering and Technology, Dhaka-1000, Bangladesh

*Corresponding author: toufiquehasan@me.buet.ac.bd

**Abstract**

Converging-diverging micro nozzle is fundamentally intended for flow acceleration through the generation of kinetic energy for the advanced micro-propulsion systems. Such supersonic micro nozzles have significant applications in the launching, propulsion and rapid directional control of the micro-satellites for a better maneuver. Micro scale analysis of such flow devices is extended to the rarefied flow regime. Present study has addressed thrust vectoring in a planar converging-diverging supersonic micro nozzle by the "bypass mass injection" technique. Direct Simulation Monte Carlo (DSMC) method has been used for numerical modelling. Primary focus is given on the thrust vectoring control of the micro nozzle with a throat height of 20 μm and an expansion ratio of 1.7. For the secondary injection, a rectangular channel of 5 different bypass widths (2-12 μm) is considered for two different outlet pressures ($P_{out}$ = 10 kPa and 40 kPa) while keeping the inlet pressure ($P_{in}$) and temperature ($T_{in}$) fixed at 1 atm and 300 k respectively. The physical behavior of the micro nozzle is acknowledged through the analysis of Mach, pressure, temperature and density contours. Numerical results reveal that the secondary flow injection is adapted into the primary flow through the formation of a pressure bump in the diverging section. Moreover, the total mass flow rate, secondary flow percentage, thrust force, the thrust coefficient and specific impulse increase with the bypass channel width. A change in thrust direction is obtained which in turn produces a considerable vectoring effect in the supersonic micro nozzle. The vectoring angle for $P_{out}$ = 40 kPa peaks at 6 μm bypass channel whereas a gradual increase of the vectoring angle with the bypass channel width is observed for $P_{out}$ = 10 kPa.

## 1. Introduction

Table 1. Satellite classification based on their mass.

| Group name | Mini satellite | Micro satellite | Nano satellite | Pico satellite | Femto satellite |
|---|---|---|---|---|---|
| Mass (kg) | 100 – 500 | 10 – 100 | 1 – 10 | 0.1 – 1 | < 0.1 |

Thrust vectoring is a mechanism for direction control of space crafts in three dimensions that enables the spacecraft to manipulate the thrust direction for altitude control. Thrust Vectoring Nozzle (TVN) technologies are designed to affect the aircraft or spacecraft dynamics by generating lateral components of thrust in the nozzle flow to control the angular velocity, altitude etc. At low speeds or high angles of attack, where the conventional aerodynamic control surfaces lose their effectiveness, thrust vectoring significantly assists in the maneuverability. If the thrust is at a different angle in the exit plane than the nozzle centerline direction, it assists in taking rapid turns along with the momentary landing and take-off capabilities. This reduces the dependency of the vehicles on the aerodynamic control surfaces for an effective flight control and maneuvering. Currently, Thrust Vector Control (TVC) is achieved by a complex arrangement of mechanical actuators that are capable of enabling flow deflection through modification of the nozzle geometry. Researchers have been studying the possibility of redirecting the thrust without the use of any mechanical actuators, rather, by actively manipulating the nozzle flow field. This concept is known as Fluidic Thrust Vectoring (FTV). This can be implemented by manipulating the main nozzle stream through a secondary injection from an external jet. There are two different ways for FTV (a) gas injection from a secondary source and (b) bypass mass injection.

In spite of having great advantages, gas injection from a secondary source is not self-reliant. Rather it requires additional flow arrangements. Starting from different types of flow control mechanism and secondary source settings, secondary injection calls upon additional design complexities along with the associated cost and maintenance efforts. These drawbacks can be overcome by bypass mass injection technique that doesn't require supplementary flow arrangement. Thus, it is easy to install even in those aerodynamic systems that weren't facilitated with any kind of vectoring mechanism initially. Moreover, bypass mass injection technique involves relatively inexpensive lightweight systems and has no moving parts. Thus its implementation is possible with the minimum penalty of the aerodynamic design. As a result of these advantages, a number of FTV techniques have been investigated in the recent years in continuum scale, such as shock vector control [1], throat shifting [2], counter flow methods [3] etc.

Deere *et al.* [4] considered a 2D nozzle for fluidic injection for flow separation control in order to associate vectoring effect with throat shifting mechanism. Substantial vectoring effect was found in their study for the highest opposition angle to the primary flow. Another vectoring mechanism was developed at NASA Langley research Center for a dual throat nozzle. Flamm *et al.* [5] reported the highest vectoring efficiency for this kind of nozzle with better thrust efficiency. This sonic plane skewing methodology implies the upstream secondary injection ends up in better vectoring effect. An extension of this study, reported by Flamm *et al.* [6], adapted the previous methodology for an axis symmetric vectored nozzle that reports the various aspects of thrust vectoring for different pressure ratios and mass injection percentage. A decrease of the discharge coefficient was observed in their analysis. Later on Shin *et al.* [7] conceptualized another dual throat thrust vectoring approach for different flow expansion characteristics. They reported that the vectoring angle increased with the secondary injection percentage and eventually got saturated to a certain level beyond which no significant change in vectoring effect was observed for higher mass injection. These studies share a common attribute of single point injection. Further development in vectoring effect can be obtained from multi-point secondary injection with mutual comparison between these two techniques. Waithe and Deere [8] pioneered a single and multiple port bypass mass injection 2D thrust vectoring nozzle for both numerical and experimental investigations and compared the effectiveness for different nozzle pressure ratio (NPR). Their results indicate that the multi-port injection results in better vectoring effect at lower pressure ratios. Similar methodology was considered by Wang *et al*. [9] where a dual throat bypass mass injection thrust vectoring phenomenon was studied for investigating a better flow adaptive capability and an increase of vectoring angle with NPR was observed. These flow manipulation techniques provide a well-developed conceptual understanding of flow physics along with the other vectoring characteristics. A different approach for thrust vectoring can be mechanical vectoring with modified nozzle geometry as reported by Kostić *et al.* [10] for a converging-diverging nozzle. They positioned a flow obstacle at the nozzle exit to deflect the flow. Their results are indicative of change in shock behavior at the diverging section due to the presence of the obstacle. Behavioral change in shock induces the vectoring effect. Such a shock controlled vectoring mechanism was reported by Li *et al.* [11]. They developed a linear relation between the thrust pitching moment and thrust vectoring angle by regulating the shock vector.

These studies are for continuum scale analysis. However, such effort is rarely found in open literature to characterize the flow structure and examine the thrust vectoring effect in case of supersonic micro nozzles. Though some studies have addressed the internal flow phenomenon analysis in micro geometries such as flow in a micro channel/nozzle in rarefied regime. The significant behavioral change between the actual flow characterization and the flow pattern predicted by the continuum scale methodology in case of micro scale analysis is reported by Xu and Shu [12]. Their study emphasizes on the molecular approach for the micro scale analysis due to continuum breakdown. Later on Akhlaghi *et al*. [13] considered this molecular flow for a constant cross section micro channel. They reported the mixing of subsonic and supersonic zones for the micro channel. Wall thermal characteristic also played in formation of shock zone in their study. Ebrahimi and Roohi [14] elevated the previous micro channel design by considering a flow in a diverging channel. Their study demonstrated a better mean gas velocity development for increasing diverging angle. All these works contributed to the deep inspiration for flow characteristics analysis in micro nozzles through molecular approach. An insightful understanding of such features came through by the experimental study of Hao *et al*. [15]. They performed their study for micro propulsion in a rectangular converging-diverging nozzle. Their study suggests that the molecular scale behavior prevents the occurrence of the shock wave due to higher viscous effect. This study was extended by Saadati and Roohi [16] for different working fluids and reported the similar generic behavior. A bimodal behavior of the velocity and temperature was observed in their study at the throat for micro nozzle that disappeared as the nozzle was scaled down to nano level. Darbandi and Roohi [17] carried out an unstructured grid based micro nozzle investigation and analyzed the effect of particle surface collision scheme variation. They reported a boundary layer separation phenomenon with reverse flow. Liu *et al*. [18] modified the DSMC method by considering a hybrid approach, continuum at inlet and rarefied at exit, with a close agreement with the reference data. Furthermore, instead of a single nozzle, Sebastião and Santos [19] considered an array of micro nozzles and analyzed the effect of surface curvature. They reported a weak dependency of nozzle performance with surface thermal load.

One of the major technical challenges in the field of space craft miniaturization is designing a suitable maneuvering system. Thrust vectoring with bypass mass injection can play a leading role in this regard. With increasing applications of Micro/Nano/Pico satellites in aerospace and other fields, micro scale thrust vectoring has the potential to become one of the most prominent maneuvering techniques for flight control mechanism of these small scale aerospace systems as given in Table 1. The vectored micro nozzles can provide an inclined thrust to the micro aerospace systems that generates sufficient torque to correct their trajectory which may have experienced an orbital disturbance caused by the drag influences and solar radiation pressure. In spite of having such a substantial amount of successful MEMS applications in aerospace industry, no such exertion has been made for a conceptual flow structure understanding of the micro scale thrust vectoring nozzles. Thus, there has been a notable research gap that is to be acknowledged in order to comprehend the physical aspects of the vectored micro nozzle flows. Present study has addressed this research gap by considering a thrust vectoring nozzle with bypass mass injection for analyzing a steady state rarefied gas flow. To the best of authors' knowledge, this is the first time thrust vectoring effect for supersonic micro nozzle in rarefied regime is being acknowledged. Computational domain is a planar converging-diverging supersonic micro nozzle with throat height of 20 μm and an expansion ratio of 1.7. Parametric investigations were carried out for variable width (2, 6, 8, 10 and 12 μm) of the bypass channel that has a perpendicular opening in the diverging section of the nozzle. The nozzle's vectoring performance is evaluated through calculating the total mass flow rate, secondary flow percentage, thrust force, thrust vectoring angle, thrust coefficient and specific impulse. The findings of the present study give new insights for thrust vectoring of a supersonic micro nozzle.

## 2. Model description

Table 2. Dimensions of the Geometry.

| $H_{in}$ | $H_t$ | $H_{exit}$ | $H_{out}$ | $H'_{out}$ | h | $L_{conv}$ | $L_{div}$ | $L_{out}$ | $L'_{out}$ |
|---|---|---|---|---|---|---|---|---|---|
| 68 | 20 | 34 | 50 | 100 | 0 - 12 | 50 | 95 | 55 | 110 |

Table 3. Molecular properties of Nitrogen gas.

| Property | Symbol | Unit | Magnitude |
|---|---|---|---|
| Diameter | $d_p$ | m | $4.17 \times 10^{-10}$ |
| Molecular mass | $m_p$ | Kg | $4.65 \times 10^{-26}$ |
| Viscosity-temperature index | $\omega$ | - | 0.74 |
| Rotational degree of freedom | $DOF_{rot}$ | - | 2 |

Figure 1 shows the schematic diagram of the present computational model in Cartesian coordinate system with respective parameters. The x-axis is aligned with the stream-wise direction having the origin at the mid-plane of the inlet. Two different computational domains are considered for the present numerical study.

i. Baseline nozzle: planar converging-diverging (CD) micro nozzle that is shown in Fig. 1 (a) with no bypass section.
ii. Vectored nozzle: a rectangular bypass channel of variable height is added to the baseline nozzle for inducing the secondary flow that opens into the diverging section at right angle as shown in Fig. 1 (b).

Mid-plane of the bypass channel starts and ends at a distance of 10 μm and 120 μm respectively from the inlet. The dimensions of the geometry are given in Table 2. All the dimensions are in μm. As the figure indicates, a specified Pressure ($P_{in}$ = 101.325 kPa) and Temperature ($T_{in}$ = 300 K) are maintained at the inlet considering Nitrogen as the working fluid. At nozzle outlet, two different outlet pressures ($P_{out}$ = 10 kPa and 40 kPa) are applied. For each of the outlet pressures, five different bypass channel dimensions (2, 6, 8, 10 and 12 μm) are taken into consideration. To simulate the particle reflection from the walls of the micro nozzle and the bypass channel, temperature based Maxwellian distribution with unit energy accommodation coefficient and tangential momentum, alternately known as diffusive wall boundary condition, is considered for a fixed wall temperature ($T_{wall}$) of 300 K. Table 3 summarizes the molecular properties of the flowing fluid. The performance of the vectored nozzle is evaluated by the following performance parameters.

Total mass flow rate,
$$m_T = \int_{A_e} \rho \overline{U} . d\overline{A_e} \qquad (1)$$

Secondary flow percentage,
$$f(\%) = \frac{m_S}{m_T} \times 100\% \qquad (2)$$

Thrust force,
$$\overline{F_T} = \int_{A_e} \overline{U}\left(\rho \overline{U}.d\overline{A}\right) + \int_{A_e} \left(P_e - P_{out}\right) d\overline{A} \qquad (3)$$

Thrust coefficient,
$$C_f = \frac{F_T}{m_T \sqrt{\frac{2\gamma RT_0}{\gamma-1}\left\{1-\left(\frac{P_e}{P_0}\right)^{\gamma-1/\gamma}\right\}}} \qquad (4)$$

Thrust vectoring angle,
$$\beta = tan^{-1}\left(\frac{F_N}{F_A}\right) \qquad (5)$$

Specific impulse,
$$I_{sp} = \frac{F_T}{m_T g} \qquad (6)$$

3. **Numerical Procedure**

3.1 DSMC method

Present study has been carried out for Nitrogen ($N_2$) gas in the rarefied condition. Knudsen number ($Kn$) signifies the degree of rarefaction and is indicated by

$$Kn = \frac{\lambda}{H_t} \qquad (7)$$

Lower limit of the Knudsen number implies that the flowing fluid is in the thermodynamic equilibrium and can be comprehended by the continuum approach. As the Knudsen number rises, the intermolecular collision becomes less frequent due to higher mean free path ($\lambda$) to characteristic length ratio and the local flow equilibrium assumption becomes invalid causing a continuum breakdown. In this situation the non-equilibrium molecular behavior plays the dominating role to characterize the flow behavior. Such a behavior is encountered by a micro/ nano nozzle that undergoes the transitional flow behavior, extended from continuum to rarefied regime. To acknowledge this molecular behavior for an exhaustive understanding of the flow physics in the vectored micro nozzle, current study has considered the particle based Direct Simulation Monte Carlo (DSMC) method for the present analysis.

Being a stochastic atomistic method for molecular flow modelling of the rarefied gas, DSMC method is founded on the kinetic theory of gases and statistical mechanics combined with particle dynamics, pioneered by Bird [20], to numerically solve the Boltzmann equation. The phenomenological description of the DSMC method underlies the fact of replacing a large cluster of real molecules by fictitious molecules/ particles known as dsmc particles through a statistical exposition within the applicability range of accuracy constraint for a better convergence speed. Such a molecular motion is described by the velocity distribution function in physical space and the corresponding velocity coordinate in terms of probabilities with normalized distribution. Boltzmann equation is an integral-differential equation to unravel the aforementioned molecular trajectories and momentum for non-equilibrium thermodynamic state and is given by

$$\frac{\partial (nf)}{\partial t} + \vec{C}.\vec{\nabla}_x (nf) + \frac{\vec{F}}{m}.\vec{\nabla}_c (nf) = \int n^2 (f'f'_z - ff_z) gbdbd\varepsilon d\vec{Z} \qquad (8)$$

Solution of the Boltzmann equation requires to split the whole computational domain into a number of grids. Binary and reverse collisions, associated with the particle motion are then simulated independently in each of the grids on a probabilistic basis. For the present study, no time counter collision partner selection model is used along with the Variable Hard Sphere (VHS) binary collision model for diatomic Nitrogen ($N_2$) gas. High fidelity binary collisions between the dsmc particles and their sampling largely depends the particles' time of residency inside a grid. Particles must occupy a grid for a certain time, large enough to ensure proper number of collisions followed by probabilistic energy redistribution. Such a stipulation can be fulfilled by taking the simulation time step to be less than or equal to one-tenth of the mean collision time and also ensuring that the grid size is less than or equal to one-third of the molecular mean free path. An indexing array authorizes the molecules pair selection for binary collisions within the grid through cross referencing of the simulated molecules. While colliding with the walls, the molecular velocity is attributed to the reflected particles according to the Maxwellian distribution based on wall temperature.

Present study has executed this mathematical modelling in the open source CFD software openFOAM in the dsmcFoam+ solver [21]. The simulation is initiated with predefined initial state preparation according to the particle specification and uniform spatial distribution over the whole computational domain. Particle collisions are then modelled by the solver, based on the intermolecular potential. As the flow is considered to be non-reactive, the corresponding collisions are facilitated with rotational, translational and vibrational energy transfer only in conformity with the particles' degrees of freedom. Each time step contributes in i) moving the dsmc particles and considering boundary interaction; ii) cross reference based particle indexing; iii) sorting for particle-particle interaction and iv) particle information sampling. The macroscopic properties result from particle sampling. Being a non-transient solution, present simulation is first allowed to reach its steady state and then the time averaging scheme was employed to the latest time step. To meet the necessary requirements for time-step, mesh cell size and no of DSMC particle per cell (PPC), guidelines specified by Saadati and Roohi [16] were strictly adhered to. For the present work, the domain was subdivided into 150 × 150 mesh size in the main nozzle body with 10 times refined mesh in the bypass section and its projection on the main nozzle. $4 \times 10^{-11}$ s time-step size with greater or equal to 12 PPC is considered to reduce the statistical error. The simulation was first allowed to reach the steady state and then the time averaging scheme was employed to the latest step for a sample size of $36 \times 10^6$.

3.2 Pressure boundary condition

According to the mathematical formulation of Wang and Li [22] 1D characteristics theory is applied for the implementation of pressure boundary condition. For a running wave, a change in velocity is formulated as

$$du = -a \frac{d\rho}{\rho} \tag{9}$$

where, $a$ is the sound speed and $\rho$ is gas density. Adopting such formulation in a boundary cell yields

$$(\rho_{out})_i - (\rho)_i = \frac{P_{out} - P_i}{a_i^2} \tag{10}$$

Ideal gas law then gives the temperature

$$(T_{out})_i = \frac{P_{out}}{(\rho_{out})_i R} \tag{11}$$

where, *i* is the boundary node index adjacent to the outlet and *P* is the pressure. The inlet velocity is obtained by interior domain extrapolation with density calculation from ideal gas law.

$$(u_{in})_i - (u)_j = \frac{P_{in} - P_j}{\rho_j a_j} \tag{12}$$

$$(\rho_{in})_j = \frac{P_{in}}{T_{in} R} \tag{13}$$

where, *j* is the inlet boundary node index.

### 3.3 Particle distribution

DSMC algorithm incorporates the particle in cell method for molecule distribution based on collision probabilities. As the molecules continuously cross the cell boundary, the cell volume will always capture a variable number of molecules at a given time. This variability can be calculated from the following Poisson's distribution with the standard deviation $\sigma(X) = \sqrt{nV}$.

$$P(X = N) = \frac{(nV)^N e^{-nV}}{N!} \tag{14}$$

*X* denotes a random variable signifying the number of molecules in a volume *V* having an average number density *n*. Such a cell occupancy of the particles results in a statistical velocity distribution function, *f* that is defined as

$$\iiint_{-\infty}^{+\infty} f(t,x,y,z,u,v,w) du dv dw = 1 \tag{15}$$

the differential velocity changes in *x*, *y* and *z*-directions are defined as *du*, *dv* and *dw*.

### 3.4 Intermolecular interactions and particle sampling

DSMC simulation requires a physical model to exemplify the binary interactions between the molecules. Molecules and neutral atoms undergo two types of intermolecular forces.

i. Pauli repulsion: resulted from electron orbitals overlapping at short distance.
ii. Weak attraction: led by dipole-dipole interrelationship.

Present dsmc simulation has considered an attractive repulsive force based molecular flow modeling due to substantial temperature alteration along the flow path. A two-step particle interaction model concerning the spherically symmetric molecular collision is employed based on the collision cross section ($\sigma_T$) and defining the scattering law. Scattering law describes the post collisional molecular velocity distribution dependency on the pre collisional velocity Molecular interaction has been implemented by considering the variable hard sphere (VHS) model due its' precise estimation of the actual intermolecular potentials within the acceptable spectrum. VHS model abides by the classical scattering law with relative collision energy dependent diameter $d$ as follows

$$\sigma_T = \pi d^2 = c\left(m_R g^2/2\right)^{-\alpha} \tag{16}$$

where $c$ and $\alpha$ are constants.

The deflection angle for the VHS model and mean free path are given by

$$\chi = 2\cos^{-1}(b/d) \tag{17}$$

$$\lambda = \frac{2\upsilon}{15}(7-2\omega)(5-2\omega)\left(2\pi R_g T\right)^{-0.5} \tag{18}$$

where $b$ is center to center distance of the two colliding molecules and $\upsilon$ is the kinematic viscosity. The macroscopic flow properties, both surface averaged and volume averaged, are derived from particle sampling by time averaging. Present study has employed flux-based time averaging for post-processing after ensuring that the steady state has been reached. While averaging, the mean flow rate of the velocity dependent gas property $h(\vec{v})$ in $x$ and $y$-directions are formulated as

$$\rho v_x h(\vec{v}) = \Sigma\left[mv_x h(\vec{v})\right]_{x+} + \Sigma\left[mv_x h(\vec{v})\right]_{x-} \tag{19}$$

$$\rho v_y h(\vec{v}) = \Sigma\left[mv_y h(\vec{v})\right]_{y+} + \Sigma\left[mv_y h(\vec{v})\right]_{y-} \tag{20}$$

3.5 Validation

To ensure the solver accuracy and proper physical modelling of the numerical procedures, validation processes are executed. The validation process evaluates whether the computational procedures are implemented accurately or not by examining the results from the current solution and comparing them with the reference data, available in open literature. For validating the present numerical modelling in dsmcFoam+ solver, a convergent-divergent micro nozzle with a rectangular cross section and a throat size of 20 μm as shown in Fig. 1 (a) was considered and validated against the experimental study of Hao *et al.* [15]. The dsmc code is verified for the similar boundary configurations as in the present case. Figure 2 shows the variation of the nozzle mass flow rate for different outlet pressures, obtained from the present computational model and that of Hao *et al.* [15]. Such comparison reveals a suitable agreement between these two models. To ensure further accuracy of the present numerical model, it was also validated against the numerical study of Saadati and Roohi [16]. Figures 3 and 4 show the comparison of Mach and temperature distribution along the centerline and nozzle exit respectively for $P_{out}$ = 10 kPa. Furthermore, Fig. 5 shows the comparison between the average exit velocity for different outlet pressures. Table 4 includes the comparison of the mass flow rate and specific impulse with the reference data. To make a qualitative comparison, Mach and pressure contours for $P_{out}$ = 10 kPa, obtained from the present model is also compared with that of Saadati and Roohi [16] and is shown in Fig. 6. Such comparison ensures sufficient accuracy for the subsequent studies.

Table 4. Comparison of the mass Flow Rate and specific impulse between present model and Saadati and Roohi [16]

| $P_{out}$ (kPa) | Mass flow rate (g/s) | | % Error | Specific Impulse (s) | | % Error |
|---|---|---|---|---|---|---|
| | Present Model | Saadati and Roohi [16] | | Present Model | Saadati and Roohi [16] | |
| 5 | 4.10 | 4.02 | 2.00 | 42.51 | 43.14 | 1.46 |
| 10 | 4.14 | 4.17 | 0.72 | 42.22 | 42.45 | 0.54 |
| 20 | 4.15 | 4.20 | 1.19 | 38.54 | 38.57 | 0.08 |
| 30 | 4.22 | 4.31 | 2.09 | 30.13 | 30.13 | 0.00 |
| 70 | 3.87 | 3.92 | 1.28 | 14.27 | 14.39 | 0.83 |

## 4. Results and Discussion

Potential expenditure reduction and flight enhancement of miniaturized space crafts has made thrust vectoring a distinguished phenomenon for aerodynamic study. The intrinsic aero physical design of the vectored nozzle and its' behavior under rarefied flow condition has been considered in the present numerical study for a speculative understanding of the flow physics. Present parametric study has considered different geometric structures of the vectored nozzle by varying the bypass section width for two different outlet pressures ($P_{out}$ = 10 kPa and 40 kPa). This approach opens a new window for aerodynamic applications of micro nozzles in rarefied condition. The results are presented in both qualitative and quantitative manners by considering Mach, pressure, temperature and density contours with different performance parameters.

### 4.1 Mach number distribution

Figures 7 and 8 show the variation of Mach distribution for different width of the bypass channel at $P_{out}$ = 10 kPa and 40 kPa respectively. On analyzing the Mach contours in Figs. 7 and 8, downstream supersonic flow is observed for all the cases. However, this supersonic region is being restricted to a smaller area with the increase of the bypass channel width as the Mach development in the diverging section is being hindered. As a result, there is a reduction of supersonic structure in the recessed cavity of the nozzle. For baseline cases, as shown in Figs. 7 (a) and 8 (a), it can be seen that sonic point ($Ma$ = 1) is achieved slightly downstream of the throat and it gradually moves towards nozzle exit as the bypass section widens. For an ideal inviscid flow, based on the area velocity relationship as given by Eqn. (21) occurrence of the sonic point is expected to be at the throat where the isentropically expanding flow shifts from the subsonic to the supersonic region.

$$\frac{dA}{A} = (M^2 - 1)\frac{du}{u} \tag{21}$$

However, this equation has been proven to be invalid in micro scale due to the associated flow irreversibility that makes the entire gas expansion a non-isentropic phenomenon. A flow subjected to frictional dissipation experiences the sonic point at the location where

$$\frac{dA}{A} = \left(\frac{(\gamma-1)T_o + (\gamma+1)T_{wall}}{4T_o}\right)\left(\frac{4C_f dx}{D}\right) \tag{22}$$

Here, $\gamma$ is specific heat ratio, $T_o$ is the total temperature, $C_f$ is the friction factor and $D$ is the hydraulic diameter. This equation predicts that the flow will always experience the sonic point at the throat downstream due to the frictional irreversibility. Now, in case of micro geometries the low Reynolds number flow experiences higher frictional dissipation as it has a larger surface to volume ratio than that of the traditional macro scale converging diverging nozzles. The ratio of frictional dissipation to total energy associated with the flowing fluid having a specific enthalpy of $E$, specific heat at constant pressure $c_p$, dissipation function $\phi$, temperature $T$, kinematic viscosity $\upsilon$ and flow velocity $u$ can be written as follows:

$$\frac{\phi}{\rho\dfrac{DE}{Dt}} \sim \frac{\upsilon u^2/L^2}{c_p T u/L} = \frac{\upsilon u}{c_p T L} \tag{23}$$

Equation (23) implies stronger frictional dissipation in the internal gas flow with the decrease of the nozzle dimension ($L$). Higher surface to volume ratio contributes to dominant viscous effects that leads to distinct throat to downstream boundary layer development. Such a circumferential aspect deteriorates the temperature declination rate near the nozzle throat and results in a larger throat temperature. As a result, a greater temperature at the throat area holds back the local Mach number causing a fall off below the sonic point. Thus the sonic point is achieved downstream of the throat.

Now, as the bypass channel becomes wider, the secondary flow percentage increases. More secondary flow injection into the diverging section induces stronger frictional effects in the nozzle internal flow. Hence, the sonic point ($Ma = 1$), which is already downstream of the throat moves further and stretches the subsonic zone from slight downstream of the throat, for baseline nozzle, to the point of secondary injection for $h/H_t = 0.6$. The development of such subsonic layer is contributed by the collaborative effect of larger surface to volume ratio, temperature dependency of the gas viscosity and wall boundary augmentation. This effect is stronger for $P_{out} = 40$ kPa compared to that of $P_{out} = 10$ kPa. Because, for $P_{out} = 40$ kPa the nozzle experiences lower pressure ratio which in turns results in lower flow acceleration. For micro nozzle, the throat Reynolds number ($Re_t$) is defined based on the throat dimension as follows

$$Re_t = \frac{u_{avg} H_t}{\upsilon} \tag{24}$$

Lower pressure ratio offers a restricted degree of flow expansion to convert the pressure energy into kinetic energy which in turns causes lesser velocity ($u_{avg}$) development and consequently lower throat Reynolds number. As explained by Liu *et al.* [18], smaller Reynolds number contributes to enhanced viscous dissipation and thicker boundary layer development. So between the two outlet pressures, gas flow for $P_{out} = 10$ kPa shows better Mach development for all the dimensions of the bypass channel. Another interesting effect of the secondary injection is the skewness of the iso-Mach lines. For no bypass and $h/H_t = 0.1$, the iso-Mach lines are symmetric about the nozzle centerline. Such a symmetry gradually breaks down for higher bypass dimensions due to the interaction between the primary and the secondary flow. The iso-Mach lines at the nozzle exit has more flattening disposition with a downward skewing tendency that signifies the flow vectoring effect. Moreover, in case of micro nozzle, the dominating viscous effect leads to lower flow expansion compared to that of the conventional inviscid ideal nozzle. Thus the reported highest Mach numbers for $P_{out} = 10$ kPa ($Ma|_{max} = 1.7$) and 40 kPa ($Ma|_{max} = 1.2$) fall off by 15% and 40% respectively below the design Mach number ($Ma = 2$), calculated from the continuum based inviscid gas flow.

To have a quantitative idea about the gradual Mach development, its' variation along the nozzle centerline is illustrated in Fig. 9. For both of the outlet pressures, the baseline nozzle and the vectored nozzle with $h/H_t = 0.1$ have identical Mach profiles. This can be attributed to the lower secondary momentum flux inclusion for $h/H_t = 0.1$ that gets dissipated in the boundary layer and has trivial effect on the centerline nozzle flow. From $h/H_t = 0.3$ to beyond, the Mach profile assumes a nearly constant characteristic with a slightly decreasing tendency at the diverging section near the bypass opening. Thus a spatially independent Mach core is developed in the center region of the diverging section. Initially this Mach core occupies a smaller region that experiences a progressive bi-directional growth with the widening of the bypass channel. This bi-directional growth of the spatially independent Mach core offers its point of initiation towards the throat in one direction while extending up to the point of secondary injection on the other direction. Slightly decreasing nature of the Mach profile in this region is anchored to lower Mach levels at higher bypass width. All the cases for both of the outlet pressures end up in supersonic outlet. However, in terms of outlet pressure dependency, the Mach profile shows different distributive structure for different outlet pressures. For $P_{out} = 10$ kPa, vectored nozzle with $h/H_t = 0.5$ shows Mach stabilization at and near sonic point ($Ma = 1$). As the bypass width increases to $h/H_t = 0.6$, nozzle diverging portion experiences a subsonic flow slightly beyond the injection point. Similar generic behavior is also found for $P_{out} = 40$ kPa. In this case, Mach stabilization at and near sonic point ($Ma = 1$) occurs for $h/H_t = 0.4$. For greater widths of the bypass channel, subsonic flow regime continues till the point of secondary injection and beyond. Moreover, for $P_{out} = 10$ kPa, baseline nozzle and the vectored nozzle with $h/H_t = 0.1$ show a continuous increase of the Mach number. Whereas for $P_{out} = 40$ kPa, the Mach number reaches to a peak at $x/H_2 = 4.85$ for the aforementioned bypass widths and then decreases. Other vectoring cases for $P_{out} = 40$ kPa also report the similar kind of peak beyond the Mach stabilization region, alternately known as quasi steady Mach region, where the position of the peak is constantly shifting towards the outlet as the bypass width increases. The peak is also decreasing. This characteristic is unique to $P_{out} = 40$ kPa. Instead of showing such peak, Mach number continues to elevate beyond the quasi steady Mach region for $P_{out} = 10$ kPa.

Figure 10 illustrates the distribution of Mach number along the throat. For both of the outlet pressures Mach distribution exhibit bimodal behavior at the center region; that is, the Mach distribution encounters a dip in the vicinity of the nozzle centerline. This is a statistical pattern where the term "mode" indicates local maxima. These local maxima are always less than the sonic point ($Ma = 1$) that complies with the previous observation. The bi-modality is more prominent for $P_{out} = 40$ kPa than that of $P_{out} = 10$ kPa except for $h/H_t = 0.5$ where the plot continues the dip till the bottom wall instead of showing the second peak. Despite the changes being minimal, we can observe a gradual decrease in throat Mach values in the mean position with increasing bypass widths indicating the weakening of Mach development. Variation of Mach number along the nozzle exit is presented in Fig. 11. Both of the outlet pressures present different traits. An all-inclusive generic trait for $P_{out} = 10$ kPa is that flow is stagnant the at the top and the bottom walls due to the particle impingement and being captured by the wall. Afterwards the Mach profile undergoes a sharp rise and takes after a parabolic/ distorted parabolic profile that is maintained throughout the rest of the entire flow cross section. This instantaneous change in Mach number is found for all the bypass dimensions at $P_{out} = 10$ kPa.

Dissimilar to a regular parabolic profile, Mach distribution for $h/H_t > 0.1$ assumes a distorted profile above the nozzle mid plane with a continuously decreasing Mach number at the mean position. This distortion in the Mach distribution slightly resemblances the bimodal behavior with two peaks, initiated for $h/H_t = 0.5$ and can further be distinguished for $h/H_t = 0.6$. For all the cases, each individual Mach profile reaches to its' local maxima at the mid plane of the nozzle and the highest local Mach number is reported for the baseline nozzle. On the other hand, Mach distribution for $P_{out} = 40$ kPa starts from zero at the wall and changes gradually instead of showing a sharp rise as in case of $P_{out} = 10$ kPa. This clearly indicates the variation in velocity boundary layer development due to different flow acceleration for $P_{out} = 40$ kPa and the associated local Reynolds number development. The Mach profile assumes a normal distributive nature in this case with increasing downward skewness for higher bypass dimensions. The highest Mach number for no bypass and $h/H_t = 0.1$ is achieved at the nozzle mid plane whereas other cases show their local maxima below the mid plane. Unlike the cases of $P_{out} = 10$ kPa, no bimodal behavior is initiated for higher bypass dimensions though they also encounter a distortion of Mach profile above the mid plane. Highest local maxima for Mach number distribution is reported for $h/H_t = 0.5$.

## 4.2 Pressure distribution

Figures 12 and 13 show the variation of pressure distribution for different bypass widths at $P_{out} = 10$ kPa and 40 kPa respectively. From the figures it's very evident that the pressure distribution is symmetric with respect to the centerline of the nozzle for no bypass and $h/H_t = 0.1$ nozzle. For the baseline case this is expected as the flow configuration and thermal stratifications are symmetric. However, for $h/H_t = 0.1$ vectored nozzle, no such symmetry in geometric configuration and flow stratification exist due to the secondary injection. Now in case of $h/H_t = 0.1$ vectored nozzle, the secondary injection is so weak that the associated momentum flux is dissipated into the boundary and does not affect the primary flow. Thus the primary floor remains symmetric. As the bypass channel width increases from $h/H_t = 0.1$, a clustering of the isobaric lines near the opening of the channel is observed. For $h/H_t = 0.4$, isobaric line surrounding the secondary opening is labelled with 50 kPa for $P_{out} = 10$ kPa and is distributed over a certain zone starting from slight downstream of the throat to the bypass opening. Similar behavior is observed for $h/H_t = 0.5$ and 0.6. For $h/H_t = 0.5$, the zone is spanned up to the near throat region and labelled with 60 kPa. Likewise, for $h/H_t = 0.6$, it is labelled with 65 kPa and is extended up to the throat of the CD nozzle. This zone is a pressure stabilizing zone and is called "pressure bump" where the pressure distribution becomes almost independent of axial co-ordinate and induces a buffering effect. All these general behaviors are true for $P_{out}$ = 40 kPa as shown in Fig. 13. For the same bypass width, the stabilization zone has a greater span for $P_{out}$ = 40 kPa than that of 10 kPa. Moreover, 40 kPa shows faster movement of the stabilization zone towards the nozzle throat that indicates the weakening of pressure drop as the bypass width increases.

The centerline pressure variation is described in Fig. 14. Similar to Mach profile, both the baseline nozzle and nozzle with $h/H_t$ = 0.1 have similar characteristics for both of the outlet pressures. For $P_{out}$ = 10 kPa the pressure continues to decline up to the nozzle exit for these two cases whereas for $P_{out}$ = 40 kPa fluid pressure reaches to its lowest extremity at $x/H_2$ = 4.85, peaking point of Mach number for the corresponding cases, and then increases. For the other bypass widths, increasing pressure bumps are observed. The pressure has a slight rising tendency in this region followed by further pressure declination. Now in case of $P_{out}$ = 40 kPa, the pressure declination following the bump region first hits its lowest extremity and again rises, unlike the pressure distribution for $P_{out}$ = 10 kPa where it continues to go down beyond the bump region. The lowest extremity gets shifted towards the nozzle exit at higher bypass dimension. Moreover, the pressure bump is being elevated and fixed to higher levels with increasing bypass dimensions for both of the outlet pressures.

Figure 15 shows the pressure distribution along the nozzle throat. A noteworthy feature of the pressure profile is that, immediately adjacent to the wall, there is a slight pressure drop for all the cases. This pressure drop is facilitated by the boundary layer development near the wall. As we go further away from the wall, pressure continues to rise until it reaches to its' local maxima at the center. Thus, a pressure distribution with two local minima near the top and the bottom walls is obtained. Similar pressure profile is also reported by Tsimpoukis *et al*. [23] for free molecular flow. Almost symmetric pressure distribution is observed at the throat for all the cases which indicates negligible effect of the secondary flow injection on the throat pressure profile. In general, the pressure profiles are distinctively rising along the throat to higher values as the bypass width increases which conforms to the previous finding that the pressure drop degrades as the bypass channel widens for both of the outlet pressures. The only exceptional pressure profile is offered by $h/H_t$ = 0.6 for $P_{out}$ = 40 kPa. Above the centerline this profile shows the regular behavior, highest pressure throughout the flow area, while showing lower pressure than that of $h/H_t$ = 0.5 below the centerline. Finally, highest separation between the two adjacent pressure profiles is observed at the center for $P_{out}$ = 10 kPa whereas for $P_{out}$ = 40 kPa, the separation is smaller near the top wall which gets wider as we move forward to the bottom wall where the highest separation is observed. Conforming to the previous exception offered by the vectored nozzle with $h/H_t = 0.6$, pressure profile separation between $h/H_t = 0.5$ and $h/H_t = 0.6$ exhibits the opposite tendency.

Figure 16 shows pressure distribution along the nozzle exit for both of the outlet pressures. At the exit section, no near wall pressure minima are present. Rather for $P_{out} = 10$ kPa, all the geometric configurations of the vectored nozzle report the lowest pressure at the top and the bottom walls, followed by a sudden increase and continuing through a regular pressure variation as shown in Fig. 16 (a). The sudden pressure jump near the wall that corresponds to the sharp rise of the Mach number. Each jump leads to a higher degree of pressure variation at the bottom end of the nozzle. Maintaining the symmetric behavior, nozzle with no bypass and $h/H_t = 0.1$ bypass reaches to its peak at the center. For the other cases the variation is skewed in the downward direction due to the secondary flow inclusion with higher local maxima at a lower vertical position. On the contrary, pressure profile for $P_{out} = 40$ kPa has a flattening structure at the nozzle exit except for $h/H_t = 0.6$. Showing a greater degree of uniformity, the pressure levels are increasing with the bypass width though the hierarchy doesn't conform to the ascending order. A pressure jump is also observed at the bottom end due to the vectoring effect whereas the top portion has no such disposition. Pressure distribution for $h/H_t = 0.6$ has a rising tendency starting from the top wall and reaches to its' peak near the bottom wall of the nozzle, showing the highest pressure throughout the entire cross section.

4.3  Temperature, density and Knudsen number distribution

As mentioned in section 4.1, the Mach development gradually gets inhibited with the increasing bypass width. This very concept can clearly be comprehended by the temperature contours provided in Figs. 17 and 18 for $P_{out} = 10$ kPa and 40 kPa respectively. From the figure it can be visualized that there is an obstruction in temperature drop with the increasing bypass width, very similar to the behavior of the pressure field. Greater secondary momentum inclusion with the increasing bypass width hinders the flow expansion and thereby reduces the rate of temperature decrement. Thus greater temperature at the outlet results in lower Mach number as Mach number is inversely proportional to the square root of the temperature. To characterize the weakening of the temperature reduction, it can be seen from Fig. 18 that the contour line representing 240 K temperature gradually moves downstream with the increasing bypass widths. Similar observation can be made for the $P_{out} = 10$ kPa as shown in Fig. 17. For both of the outlet pressures, the isotherms are getting more and more skewed at the nozzle exit with a flattening disposition at higher bypass dimensions. Between the two outlet pressures, higher temperature drop is observed for $P_{out}$ = 10 kPa. For instance, the nozzle exit temperature drops to 240 K at $P_{out}$ = 40 kPa for $h/H_t$ = 0.5 whereas for the $P_{out}$ = 10 kPa, it drops to 210 K. This is in general true for all the other bypass dimensions as well. A more comprehensive understanding of the thermal stratification can be gathered by projecting the temperature variation along the centerline as shown in Fig. 19. The general features are similar to that of the pressure variation. For $P_{out}$ = 10 kPa, temperature decreases continuously for the baseline nozzle and $h/H_t$ = 0.1 nozzle. The secondary flow inclusion is characterized by the evolving temperature stabilization that can be termed as thermal quasi steadiness. The term "quasi" stands for a slightly rising tendency of the temperature in this zone. Similar steadiness is also reported for $P_{out}$ = 40 kPa followed by further temperature decrement that reaches to its' minima near the nozzle exit. This exactly resemblances the minima corresponding to the pressure distribution that constantly moves towards the nozzle exit with increasing bypass dimensions. Moreover, for the same bypass width, the quasi steady behavior is anchored to a lower temperature level for $P_{out}$ = 10 kPa than that of $P_{out}$ = 40 kPa.

Figure 20 illustrates the temperature distribution along the throat. The temperature distribution exhibits a bimodal behavior; that is, the center region is warmer than the adjacent fluid. Similar phenomenon is also observed for the rarefied Poiseuille flow and is reported by Uribe and Garcia [24]. Bimodal behavior of the temperature that classical Navier stokes model fail to capture was theoretically explained by Tij and Santos [25] using Bhatnagar-Gross-Kook model and Risso and Cordero [26] by Grad's expansion method. Upon comparing with Mach distribution, this bimodal behavior is more conspicuous in the temperature distribution than in case of Mach distribution. Highest temperature is at the wall that is fixed to 300 K and then it decreases to its' local minima above the center plane while introducing the local maxima at the center and further followed by the second local minima below the center plane for all the bypass dimensions. No general hierarchical attribute can be assigned to the temperature profiles based on the ascending bypass width. The dip at the center is more eminent for $P_{out}$ = 40 kPa than that of $P_{out}$ = 10 kPa while displaying a symmetric nature. Temperature variation along the exit plane is represented in Fig. 21 for both of the outlet pressures. This time the bimodal behavior of the temperature is absent and the behavioral phenomenon has quite similarities with the Mach number distribution at the nozzle exit. At $P_{out}$ = 10 kPa, a slight temperature jump can be noticed at the top and the bottom walls followed by a parabolic temperature variation for the baseline and $h/H_t$ = 0.1 nozzle and a distorted parabolic profile for the other cases. The distortion leads to the initiation of a slight bi-modal tendency of the temperature just same as the Mach profile at the same vertical positions. $P_{out}$ = 40 kPa also demonstrates the same generic structure as that of the Mach number with a naturally distributive nature of temperature for the first two cases followed by downward skewed profiles for the other cases. While the temperature profiles for $P_{out}$ = 10 kPa is reporting very distinctive local minima at the center, temperature profiles for $P_{out}$ = 40 kPa is showing the same minima for all the bypass dimensions but at different vertical positions.

Analysis of the density contours are presented in Figs. 22 and 23 that reveal the flow concentration phenomenon. For instance, Figs. 22 and 23 disclose that the contour line representing 0.7 kg/m³ gradually shifts towards downstream as the bypass channel widens. This clearly hints at the thwarting of the density drop. Gradual biasness of the contour lines, downstream of the injection point, can also be visualized from the figure. Moreover, the descending magnitude of the density contour lines signify less molecular congregation that comprehends a certain physical space. Thus intermolecular distance is increasing in terms of physical interval. Enhanced physical interval manifests a longer travelling path between two molecular collisions. Variation of the mean free path along the centerline quantify this flow attitude in terms of Knudsen number ($Kn$) as shown in Fig. 24. It can be seen that for the whole flow path Knudsen number lies within the range of $10^{-3} < Kn < 10^{-1}$. Thus the entire flow is in the slip regime for both of the outlet pressures that justify the choice of dsmc method to conduct the present computational study. Decreased flow density leads to greater mean collision time that allows larger mean free path ($\lambda$). Thus Knudsen number is increasing along the centerline. All the flow cases for both of the outlet pressures have similar Knudsen number ($Kn$) up to the throat. This is because the variable hard sphere mean free path is contributed by the combined effect of flow pressure and temperature. From Figs. 14 and 19 it can be seen that the flow pressure and temperature has slight or no variation along the centerline till the throat which results in an invariant $Kn$. $P_{out}$ = 10 kPa shows continuously increasing Knudsen number for no bypass and $h/H_t$ = 0.1 nozzle. The other cases show $Kn$ stabilization due to density thwarting caused by the pressure bump. Such stabilization is common for both of the outlet pressures. The stabilization zone for $Kn$ has a slight decreasing tendency that is similar to the Mach number distribution. $Kn$ variation for $P_{out}$ = 40 kPa reports a peak near the nozzle exit, continuously shifting outward as the bypass section widens, that corresponds to the Mach peak as shown in Fig. 14. These similarities between the Mach number variation and the Knudsen number variation along the nozzle centerline arises as both the mean free path and Mach number are inversely proportional to the square root of temperature.

### 4.4 Vectoring performance

The variation of the total mass flow rate and the secondary flow percentage with the bypass channel width is shown in Fig. 25 (a). The total mass flow rate increases as the bypass channel widens. This is due to higher secondary flow rate as wider bypass channel offers larger secondary flow area. The two curves representing the mass flow rate for both of the outlet pressures are quite identical indicating that they have weak dependency on the outlet pressure. However, at the beginning, mass flow rate corresponding to $P_{out}$ = 40 kPa is slightly higher than that of $P_{out}$ = 10 kPa. The hierarchy continues up to $h/H_t$ = 0.3. Beyond that the mass flow rate corresponding to $P_{out}$ = 10 kPa is higher and the highest variance is observed for $h/H_t$ = 0.6. Increasing the bypass width enhances the pressure bump that stabilizes the pressure. Thus the primary flow experiences less pressure gradient and the flow expansion gets restricted that may deteriorate the primary flow. However, the secondary flow through the bypass section gets higher exposure at the same time due to larger flow area that more than offsets the effect of pressure bump and eventually improves the total mass flow rate. Higher secondary mass flow rate at larger bypass width is reflected by the gradually increasing secondary flow percentage. For both of the outlet pressures the secondary flow percentage plots are identical, signifying that outlet pressure has trivial effect on them.

Figure 25 (b) shows the variation of the thrust force and the thrust coefficient. Thrust force is continuously increasing with the bypass width for both of the outlet pressures. This situation is mostly due to the higher total mass flow rate. Thrust force corresponding to $P_{out}$ = 10 kPa is always higher. This can be well explained by the distributive nature of the Mach contour. The thrust force is contributed by (i) momentum flux and (ii) pressure thrust where the momentum flux plays the most significant role. It's very evident from the Mach contour that for $P_{out}$ = 10 kPa, a structured supersonic zone is created earlier inside the nozzle than that of $P_{out}$ = 40 kPa for all bypass widths. Therefore, the supersonic zone exists for a longer span inside the recessed cavity of the nozzle that ends up at higher Mach value at the nozzle exit. This phenomenon induces a higher momentum development and eventually leads to larger thrust force. As the bypass width increases, Mach number at the nozzle exit decreases signifying flow restriction due to viscous dissipation. This phenomenon definitely has a detrimental effect on the thrust force as the supersonic zone inside the recessed cavity of the nozzle becomes smaller with the gradual increase of the bypass width. This negative effect is compensated by the additional mass flow rate, coming through the secondary opening which contributes to the thrust generation by the additional momentum development. As a result, the thrust force continues to increase. Similar reasoning holds for the gradual increase of thrust coefficient. For both of the outlet pressures, thrust coefficient is increasing almost linearly with higher thrust coefficient for $P_{out}$ = 10 kPa. This a direct implication of higher thrust force at $P_{out}$ = 10 kPa due to a better flow expansion. Estakhrsa *et al.* [27] also found the similar data trend for the continuum scale converging diverging nozzle. Although increasing, the thrust coefficient for vectored micro nozzle is very less compared to that of in the continuum regime (> 0.9). This is due to the combination of viscous, thermal, and rarefaction effects on the microscale flow structure that considerably affects the supersonic flow behavior in micro nozzles. As explained by Alexeenko *et al.* [28], the viscous effects will dominate the gas expansion and reduce the thrust mainly due to significant wall shear stress and thus a micro nozzle will always show less thrust coefficient than that of a conventional nozzle.

One of the most fundamental applications of the vectored micro nozzle is the flight trajectory correction of the micro/nano satellites by the maneuvering moment that is a direct function of the thrust vectoring angle. The higher the vectoring angle the greater the maneuvering moment. Vectoring angle is a measurement of how much the flow is being bended inside the cavity due to the secondary inclusion. Variation of the thrust vectoring angle is presented in Fig. 25 (c). Here two alternate behaviors are being noticed. For $P_{out}$ = 10 kPa, thrust vectoring angle is continuously increasing due to higher mass flow percentage. Higher momentum flow rate is associated with this larger secondary flow. Thus his secondary flow can easily bend the primary flow and continues to increase the thrust vector angle. Similar behavior is also reported by Estakhrsar *et al.* [27] for continuum scale analysis. Now vectoring angle for $P_{out}$ = 40 kPa has two distinctive features. Firstly, it shows higher vectoring angle than that of $P_{out}$ = 10 kPa up to $h/H_t$ = 0.4. Secondly, the thrust vectoring angle for $P_{out}$ = 40 kPa shows a peak at $h/H_t$ = 0.3 beyond which the vectoring angle decreases with the increase of bypass width. Similar peak is also reported by Wang *et al.* [9] for continuum scale thrust vectoring. This feature can be used as a design parameter for a better flight control and obtain the maximum maneuvering effect.

Figure 25 (d) shows the variation of specific impulse with bypass width. Specific impulse is a measure of how efficiently the propellant of a thruster is being utilized to generate the thrust. This is a typical parameter to compare among different flight methods for various propellants. Mathematically it is defined by the thrust generated by per unit weight flow of the fuel. From the figure, it is evident that the specific impulse is increasing with bypass dimension for both of the outlet pressures. This implies higher thrust to weight flow ratio of the vectored nozzle. Thus less propellant will be needed to achieve a certain amount of thrust. Vectored nozzle experiences higher increment rate of the thrust force than that of mass flow rate which increases the specific impulse. As specific impulse is inversely proportional to specific fuel consumption, vectored nozzle with greater bypass dimension is more fuel efficient. The outlet pressure seems to have a strong influence on specific impulse since the two curves are very distinct. $P_{out} = 10$ kPa and $P_{out} = 40$ kPa has similar mass flow rate while $P_{out} = 10$ kPa allows greater flow expansion and results in higher thrust. As a result, higher specific impulse is reported for $P_{out} = 10$ kPa than that of $P_{out} = 40$ kPa. Moreover, effective exhaust velocity ($u_e$) of the nozzle is associated with the specific impulse and can be calculated as $u_e = F_T/m_T$. Higher specific impulse corresponds to greater effective velocity. Table 5 summarizes the variation of the effective exhaust velocity with the bypass dimensions.

Table 5. Variation of the effective exhaust velocity for different bypass widths.

| Non-dimensional bypass width, ($h/H_t$) | Effective exhaust velocity, $u_e$ (m/s) | |
|---|---|---|
| | $P_{out} = 10$ kPa | $P_{out} = 40$ kPa |
| 0.0 | 471.47 | 226.61 |
| 0.1 | 476.93 | 234.16 |
| 0.3 | 481.70 | 251.22 |
| 0.4 | 487.26 | 263.45 |
| 0.5 | 491.33 | 276.59 |
| 0.6 | 492.34 | 280.36 |

5. **Conclusion**

Findings of the present computational study can be summarized as follows.
i. Secondary injection introduces pressure bump in the diverging section.
ii. Total mass flow rate through the converging-diverging nozzle increases with the bypass channel width with a weak dependency on the outlet pressure.
iii. Along with the total flow, the secondary flow percentage also increases with the bypass width.
iv. Thrust force increases continuously with the bypass channel width for both of the outlet pressures.
v. For all of the cases, thrust force corresponding to $P_{out} = 10$ kPa is greater than that of $P_{out} = 40$ kPa.

vi. For $P_{out}$ = 10 kPa, thrust vectoring angle increases continuously whereas for $P_{out}$ = 40 kPa the thrust vectoring angle peaks at $h/H_t$ = 0.3 and decreases beyond that.

vii. Higher thrust force for $P_{out}$ = 10 kPa results in higher thrust coefficient.

**Conflict of Interest**

The authors declare no conflict of interest.

**Acknowledgement**

**References**

[1] R. Deng, T. Setoguchi, and H. D. Kim, "Large eddy simulation of shock vector control using bypass flow passage," *International Journal of Heat and Fluid Flow*, vol. 62, pp. 474–481, Dec. 2016, doi: 10.1016/j.ijheatfluidflow.2016.08.011.

[2] P. J. Yagle, D. N. Miller, K. B. Ginn, and J. W. Hamstra, "Demonstration of Fluidic Throat Skewing for Thrust Vectoring in Structurally Fixed Nozzles," *Journal of Engineering for Gas Turbines and Power*, vol. 123, no. 3, pp. 502–507, Jan. 2001, doi: 10.1115/1.1361109.

[3] J. Flamm, "Experimental study of a nozzle using fluidic counterflow for thrust vectoring," in *34th AIAA/ASME/SAE/ASEE Joint Propulsion Conference and Exhibit*, American Institute of Aeronautics and Astronautics.

[4] K. Deere, B. Berrier, J. Flamm, and S. Johnson, "Computational Study of Fluidic Thrust Vectoring Using Separation Control in a Nozzle," in *21st AIAA Applied Aerodynamics Conference*, American Institute of Aeronautics and Astronautics.

[5] J. Flamm, K. Deere, M. Mason, B. Berrier, and S. Johnson, "Design Enhancements of the Two-Dimensional, Dual Throat Fluidic Thrust Vectoring Nozzle Concept," in *3rd AIAA Flow Control Conference*, American Institute of Aeronautics and Astronautics.

[6] J. Flamm, K. Deere, M. Mason, B. Berrier, and S. Johnson, "Experimental Study of an Axisymmetric Dual Throat Fluidic Thrust Vectoring Nozzle for Supersonic Aircraft Application," in *43rd AIAA/ASME/SAE/ASEE Joint Propulsion Conference & Exhibit*, American Institute of Aeronautics and Astronautics.

[7] C. S. Shin, H. D. Kim, T. Setoguchi, and S. Matsuo, "A computational study of thrust vectoring control using dual throat nozzle," *J. Therm. Sci.*, vol. 19, no. 6, pp. 486–490, Dec. 2010, doi: 10.1007/s11630-010-0413-x.

[8] K. Waithe and K. Deere, "An Experimental and Computational Investigation of Multiple Injection Ports in a Convergent-Divergent Nozzle for Fluidic Thrust Vectoring," in *21st AIAA Applied Aerodynamics Conference*, American Institute of Aeronautics and Astronautics.

[9] Y. Wang, J. Xu, S. Huang, Y. Lin, and J. Jiang, "Computational study of axisymmetric divergent bypass dual throat nozzle," *Aerospace Science and Technology*, vol. 86, pp. 177–190, Mar. 2019, doi: 10.1016/j.ast.2018.11.059.

[10] O. Kostić, Z. Stefanović, and I. Kostić, "CFD modeling of supersonic airflow generated by 2D nozzle with and without an obstacle at the exit section," *FME Transactions*, vol. 43, no. 2, pp. 107–113, 2015, doi: 10.5937/fmet1502107k.

[11] L. Li, M. Hirota, K. Ouchi, and T. Saito, "Evaluation of fluidic thrust vectoring nozzle via thrust pitching angle and thrust pitching moment," *Shock Waves*, vol. 27, no. 1, pp. 53–61, Jan. 2017, doi: 10.1007/s00193-016-0637-0.

[12] H. Xue, Q. Fan, and C. Shu, "Prediction of micro-channel flows using direct simulation Monte Carlo," *Probabilistic Engineering Mechanics*, vol. 15, no. 2, pp. 213–219, Apr. 2000, doi: 10.1016/S0266-8920(99)00023-5.

[13] H. Akhlaghi, M. Balaj, and E. Roohi, "Direct simulation Monte Carlo investigation of mixed supersonic–subsonic flow through micro-/nano-scale channels," *Physica Scripta*, vol. 88, no. 1, p. 015401, May 2013, doi: 10.1088/0031-8949/88/01/015401.

[14] A. Ebrahimi and E. Roohi, "DSMC investigation of rarefied gas flow through diverging micro and nanochannels," *Microfluid Nanofluid*, vol. 21, no. 2, p. 18, Feb. 2017, doi: 10.1007/s10404-017-1855-1.

[15] P. F. Hao, Y.-T. Ding, Z.-H. Yao, F. He, and K.-Q. Zhu, "Size effect on gas flow in micro nozzles," *J. Micromech. Microeng.*, vol. 15, no. 11, pp. 2069–2073, Sep. 2005, doi: 10.1088/0960-1317/15/11/011.

[16] S. A. Saadati and E. Roohi, "Detailed investigation of flow and thermal field in micro/nano nozzles using Simplified Bernoulli Trial (SBT) collision scheme in DSMC," *Aerospace Science and Technology*, vol. 46, pp. 236–255, Oct. 2015, doi: 10.1016/j.ast.2015.07.013.

[17] M. Darbandi and E. Roohi, "Study of subsonic–supersonic gas flow through micro/nanoscale nozzles using unstructured DSMC solver," *Microfluid Nanofluid*, vol. 10, no. 2, pp. 321–335, Feb. 2011, doi: 10.1007/s10404-010-0671-7.

[18] M. Liu, X. Zhang, G. Zhang, and Y. Chen, "Study on micronozzle flow and propulsion performance using DSMC and continuum methods," *Acta Mech Mech Sinica*, vol. 22, no. 5, pp. 409–416, Oct. 2006, doi: 10.1007/s10409-006-0020-y.

[19] I. B. Sebastião and W. F. N. Santos, "Numerical simulation of heat transfer and pressure distributions in micronozzles with surface discontinuities on the divergent contour," *Computers & Fluids*, vol. 92, pp. 125–137, Mar. 2014, doi: 10.1016/j.compfluid.2013.12.023.

[20] G. A. Bird, "Molecular Gas Dynamics and The Direct Simulation of Gas Flows," *Molecular Gas Dynamics and The Direct Simulation of Gas Flows, Clarendon Press: Oxford*, 1994.

[21] M. R. Wang and Z. X. Li, "Numerical simulations on performance of MEMS-based nozzles at moderate or low temperatures," *Microfluid Nanofluid*, vol. 1, no. 1, pp. 62–70, Nov. 2004, doi: 10.1007/s10404-004-0008-5.

[22] OpenFOAM (2009), The Open Source CFD toolbox, user guide, version 2.4, http://www.openfoam.org/.

[23] A. Tsimpoukis, N. Vasileiadis, G. Tatsios, and D. Valougeorgis, "Nonlinear oscillatory fully-developed rarefied gas flow in plane geometry," *Physics of Fluids*, vol. 31, no. 6, p. 067108, Jun. 2019, doi: 10.1063/1.5099051.

[24] F. J. Uribe and A. L. Garcia, "Burnett description for plane Poiseuille flow," *Phys. Rev. E*, vol. 60, no. 4, pp. 4063–4078, Oct. 1999, doi: 10.1103/PhysRevE.60.4063.

[25] M. Tij and A. Santos, "Perturbation analysis of a stationary nonequilibrium flow generated by an external force," *J Stat Phys*, vol. 76, no. 5, pp. 1399–1414, Sep. 1994, doi: 10.1007/BF02187068.

[26] D. Risso and P. Cordero, "Generalized hydrodynamics for a Poiseuille flow: Theory and simulations," *Phys. Rev. E*, vol. 58, no. 1, pp. 546–553, Jul. 1998, doi: 10.1103/PhysRevE.58.546.

[27] M. H. Hamedi-Estakhrsar, M. Ferlauto and H. Mahdavy-Moghaddam, 2020. "Numerical study of secondary mass flow modulation in a Bypass Dual-Throat Nozzle," *Proceedings of*

*the Institution of Mechanical Engineers, Part G: Journal of Aerospace Engineering*, p.0954410020947920, July. 2020, doi: 10.1177/0954410020947920.

[28] A. A. Alexeenko, D. A. Fedosov, S. F. Gimelshein, D. A. Levin, and R. J. Collins, "Transient heat transfer and gas flow in a MEMS-based thruster," *Journal of Microelectromechanical Systems*, vol. 15, no. 1, pp. 181–194, Feb. 2006, doi: 10.1109/JMEMS.2005.859203.

**Nomenclature**

| Symbol | Units | Description |
| --- | --- | --- |
| $A_e$ | [m$^2$] | Exit plane area |
| $C_f$ | [-] | Thrust coefficient |
| $\vec{C}$ | [m.s$^{-1}$] | Initial velocity of the colliding particle |
| $\vec{C}\,'$ | [m.s$^{-1}$] | Final velocity of the colliding particle |
| $f_z$ | [-] | Velocity distribution function of the collision partner before collision |
| $f_z'$ | [-] | Velocity distribution function of the collision partner after collision |
| $\vec{F}$ | [N] | External force |
| $F_T$ | [N] | Thrust force |
| $F_A$ | [N] | Axial thrust component |
| $F_N$ | [N] | Normal thrust component |
| $k$ | [J.K$^{-1}$] | Boltzmann constant |
| $Kn$ | [-] | Knudsen number |
| $m_T$ | [g.s$^{-1}$] | Total mass flow rate |
| $m_S$ | [g.s$^{-1}$] | Secondary mass flow rate |
| $Ma$ | [-] | Mach number |
| $n$ | [m$^{-3}$] | Number density |
| $NPR$ | [-] | Nozzle pressure ratio |
| $P$ | [Pa] | Pressure |
| $P_e$ | [Pa] | Nozzle exit pressure |
| $PPC$ | [-] | Particle per cell |
| $R_g$ | [J.kg$^{-1}$.K$^{-1}$] | Specific gas constant |
| $T$ | [K] | Temperature |
| $u$ | [m.s$^{-1}$] | Flow velocity |
| $\vec{Z}$ | [m.s$^{-1}$] | Initial velocity of the collision partner |
| $\vec{Z}\,'$ | [m.s$^{-1}$] | Final velocity of the collision partner |

| Greek letters | | |
|---|---|---|
| $\beta$ | [degree] | Vector angle |
| $\varepsilon$ | [-] | Collision parameter |
| $\gamma$ | [-] | Specific heat ratio |
| $\lambda$ | [m] | Mean free path |
| $\upsilon$ | [m$^2$.s$^{-1}$] | Kinematic viscosity |
| $\rho$ | [kg.m$^{-3}$] | Density |
| Subscripts | | |
| $w$ | | Wall |
| $in$ | | Inlet |
| $out$ | | Outlet |
| $e$ | | Exit |
| $Z$ | | Collision partner |

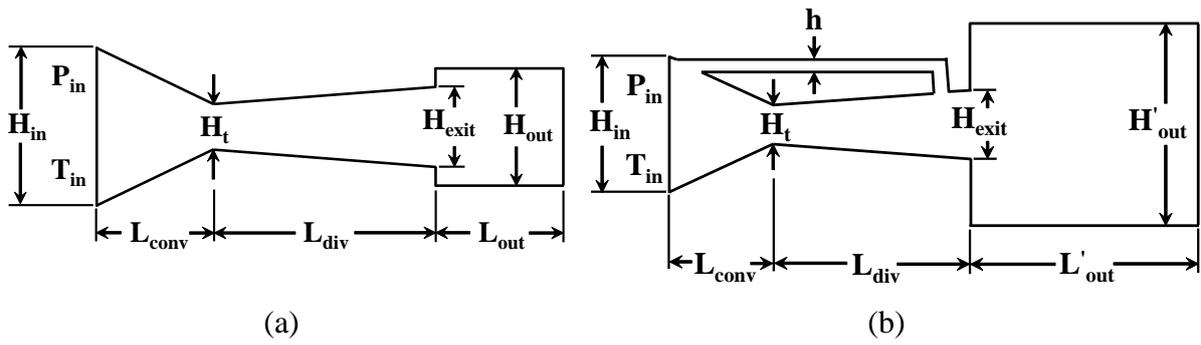

**Figure 1.** Converging-diverging (a) baseline nozzle and (b) thrust vectoring nozzle with imposed boundary conditions.

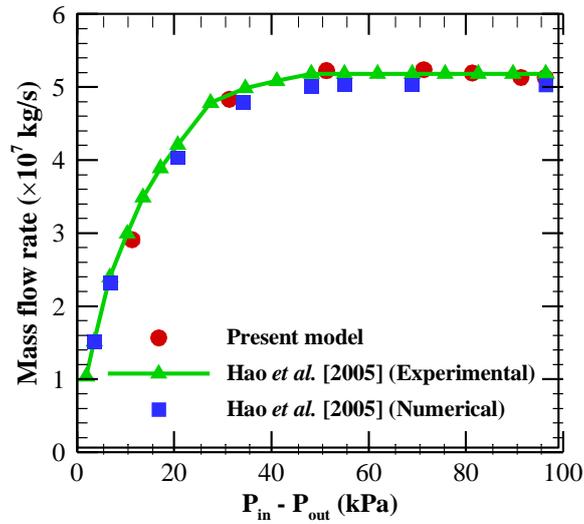

**Figure 2.** Comparison of the mass flow rate between the present model and that of Hao *et al.* [2005].

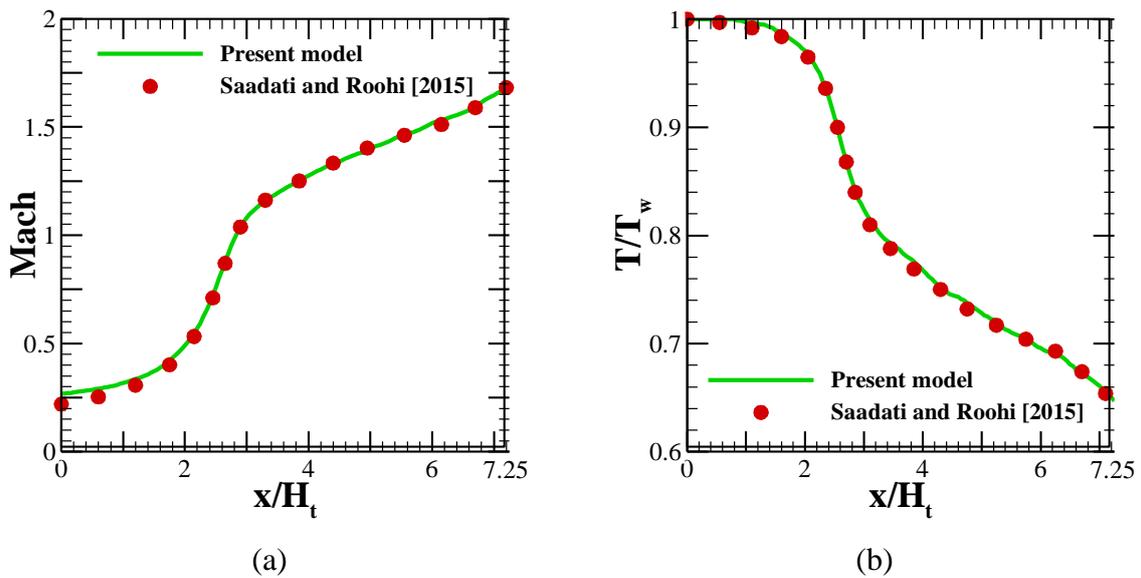

**Figure 3.** Comparison of the (a) Mach number and (b) normalized temperature distribution along the centerline for $P_{out} = 10$ kPa.

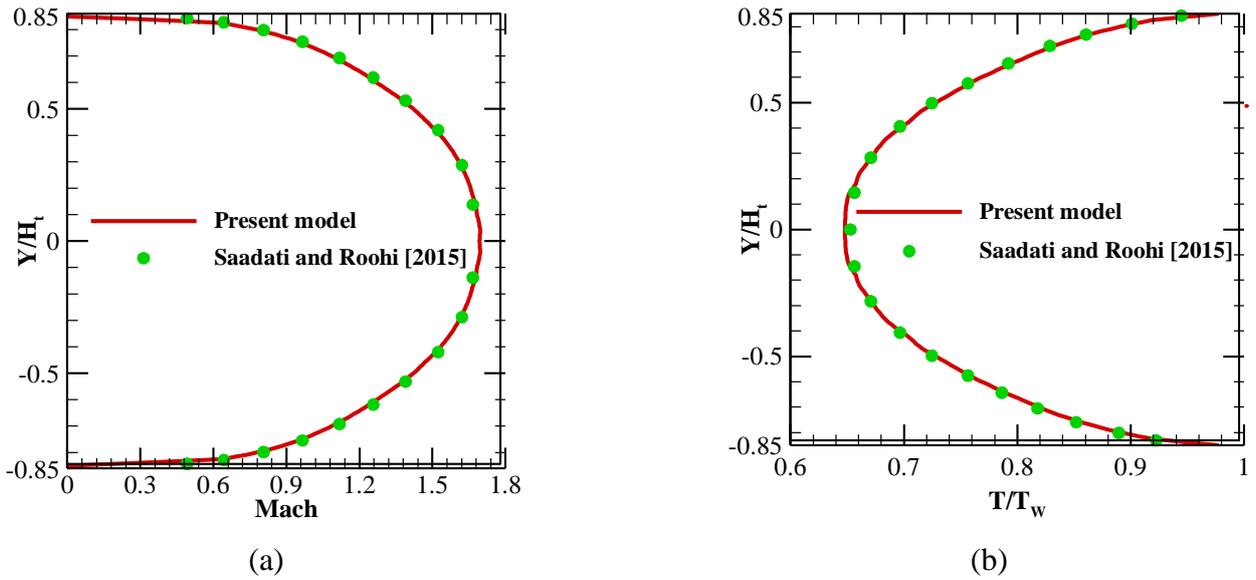

(a)                          (b)

**Figure 4.** Comparison of (a) Mach number and (b) normalized temperature distribution along the exit plane for $P_{out}$ = 10 kPa

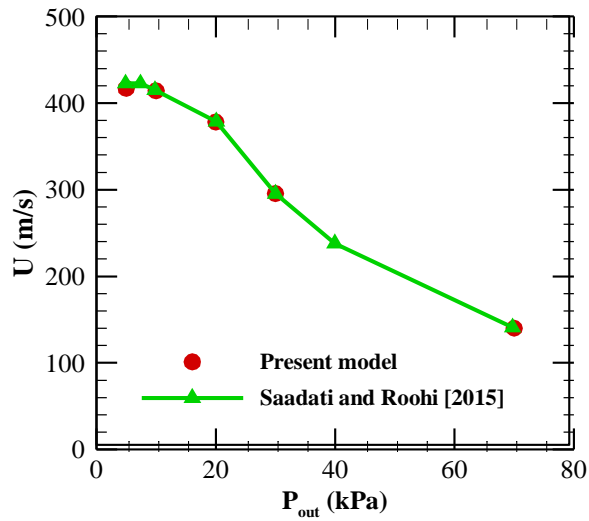

**Figure 5.** Comparison of the nozzle exit velocities at different outlet pressures.

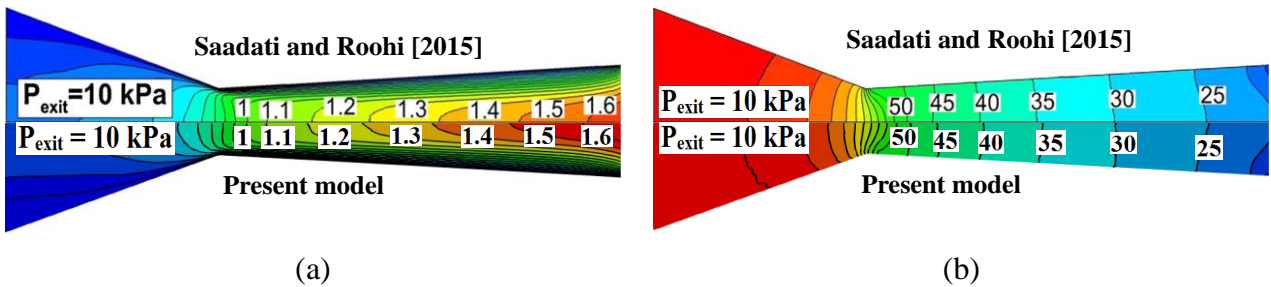

(a)                          (b)

**Figure 6.** Comparison of the (a) Mach and (b) pressure (kPa) contours at $P_{out}$ = 10 kPa

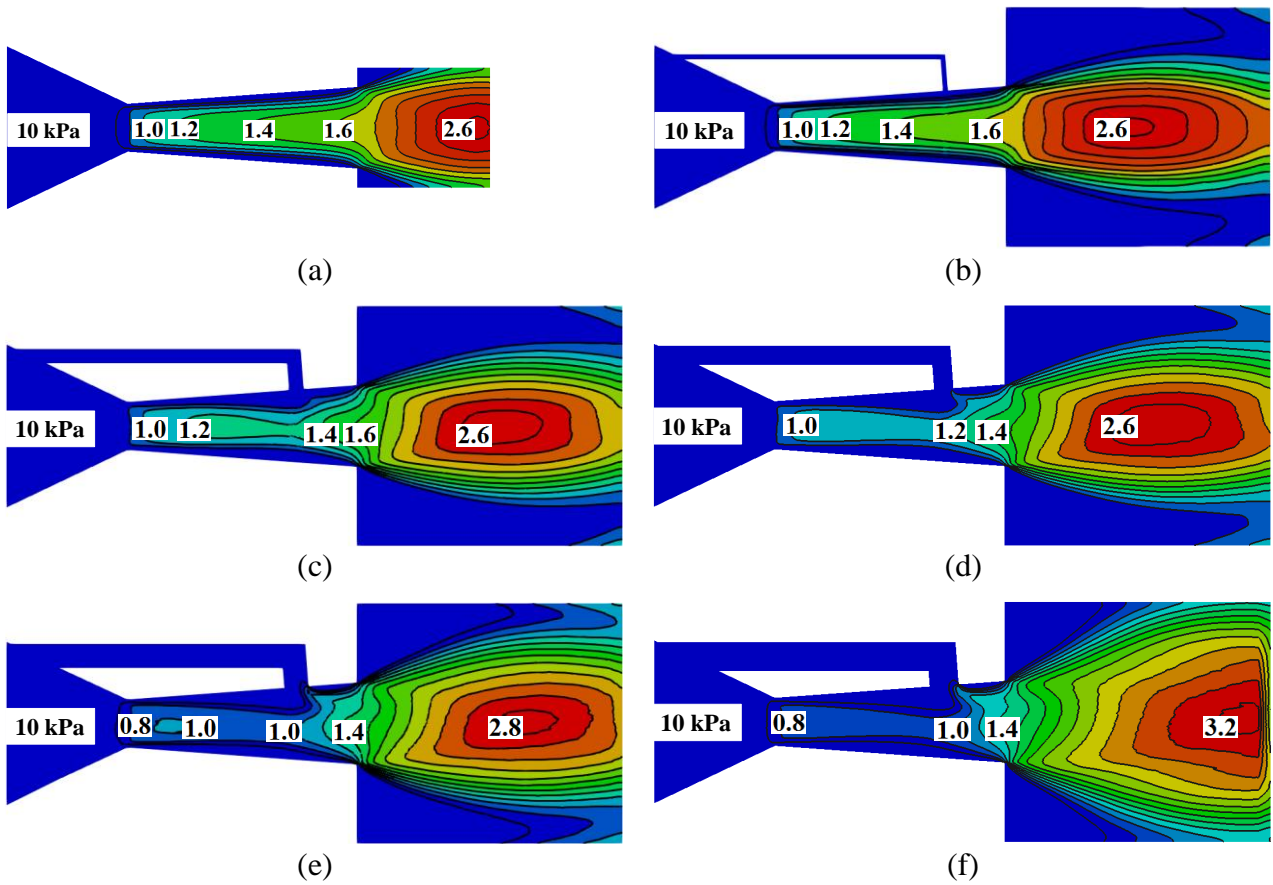

**Figure 7.** Mach contours for (a) baseline nozzle, (b) $h/H_t = 0.1$, (c) $h/H_t = 0.3$, (d) $h/H_t = 0.4$, (e) $h/H_t = 0.5$ and (f) $h/H_t = 0.6$ thrust vectoring nozzle at $P_{out} = 10$ kPa.

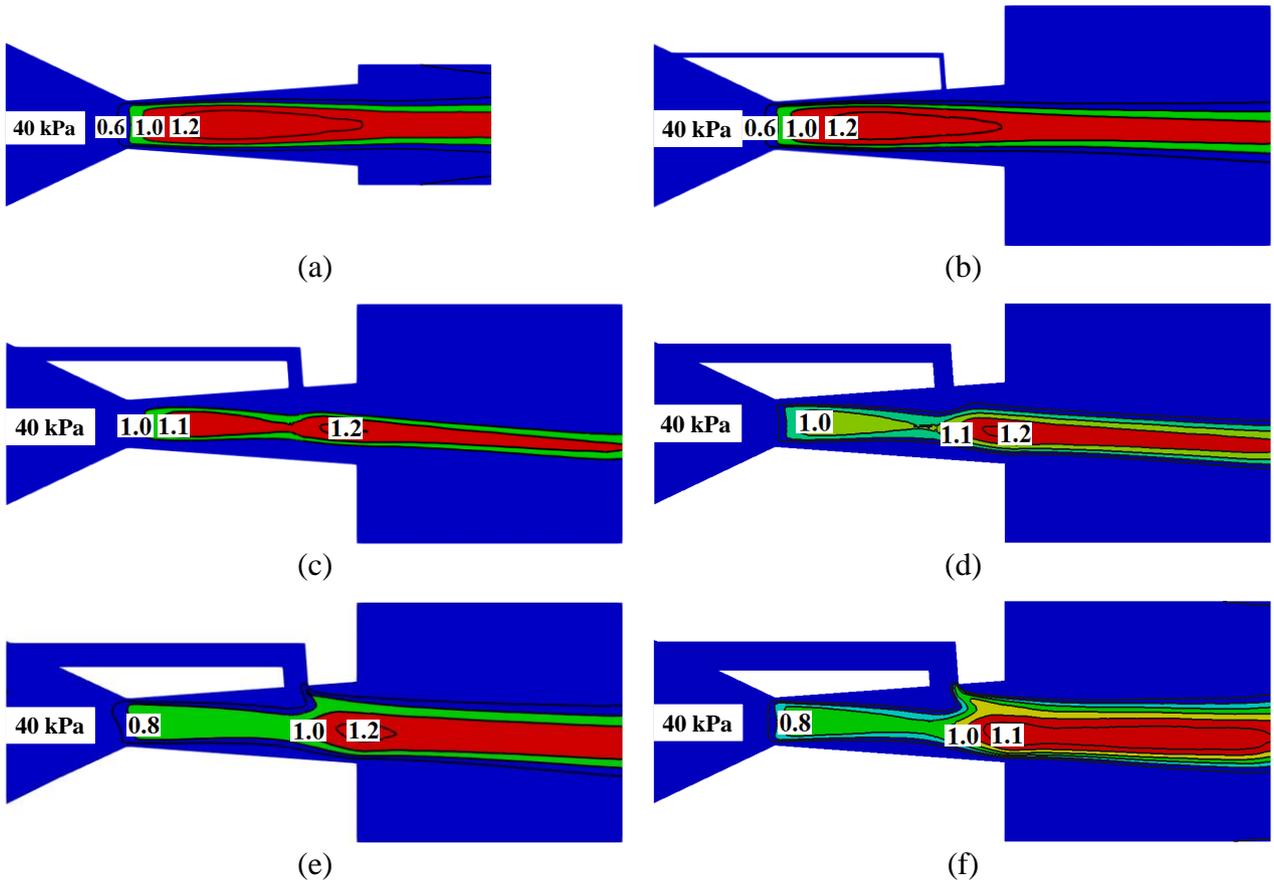

**Figure 8.** Mach contours for (a) baseline nozzle, (b) $h/H_t = 0.1$, (c) $h/H_t = 0.3$, (d) $h/H_t = 0.4$, (e) $h/H_t = 0.5$ and (f) $h/H_t = 0.6$ thrust vectoring nozzle at $P_{out} = 40$ kPa.

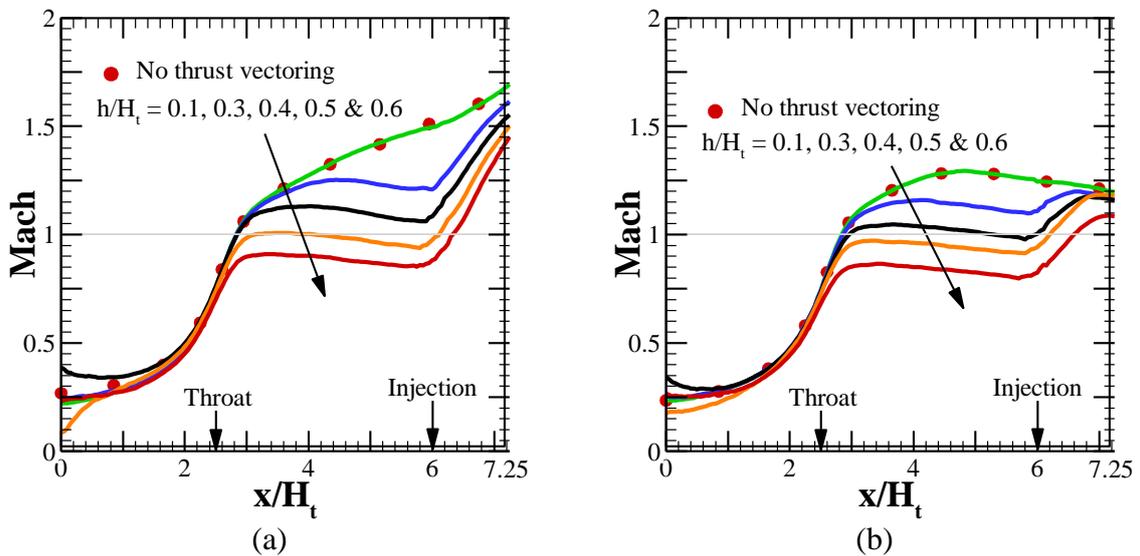

**Figure 9.** Variation of the Mach number along the centerline at (a) $P_{out} = 10$ kPa and (b) $P_{out} = 40$ kPa for different bypass channel dimensions.

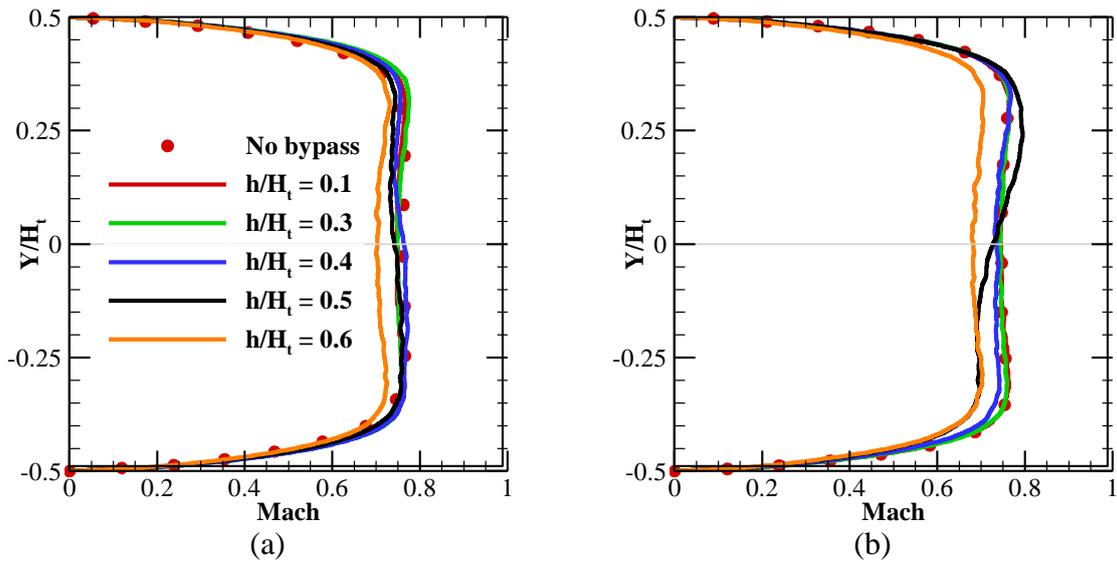

**Figure 10.** Variation of the Mach number along the throat at (a) $P_{out}$ = 10 kPa and (b) $P_{out}$ = 40 kPa for different bypass channel dimensions.

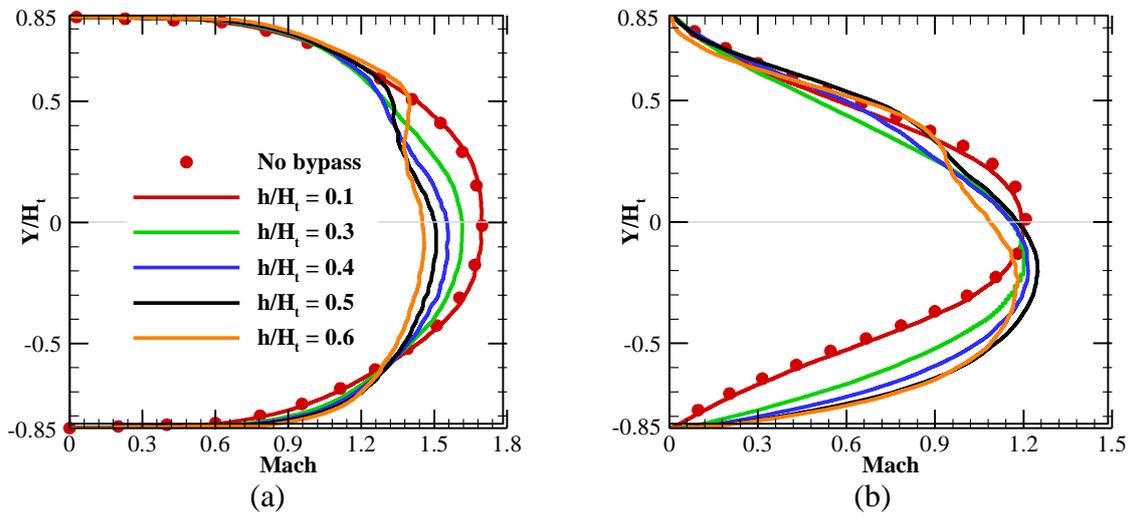

**Figure 11.** Variation of the Mach number along the exit plane at (a) $P_{out}$ = 10 kPa and (b) $P_{out}$ = 40 kPa for different bypass channel dimensions.

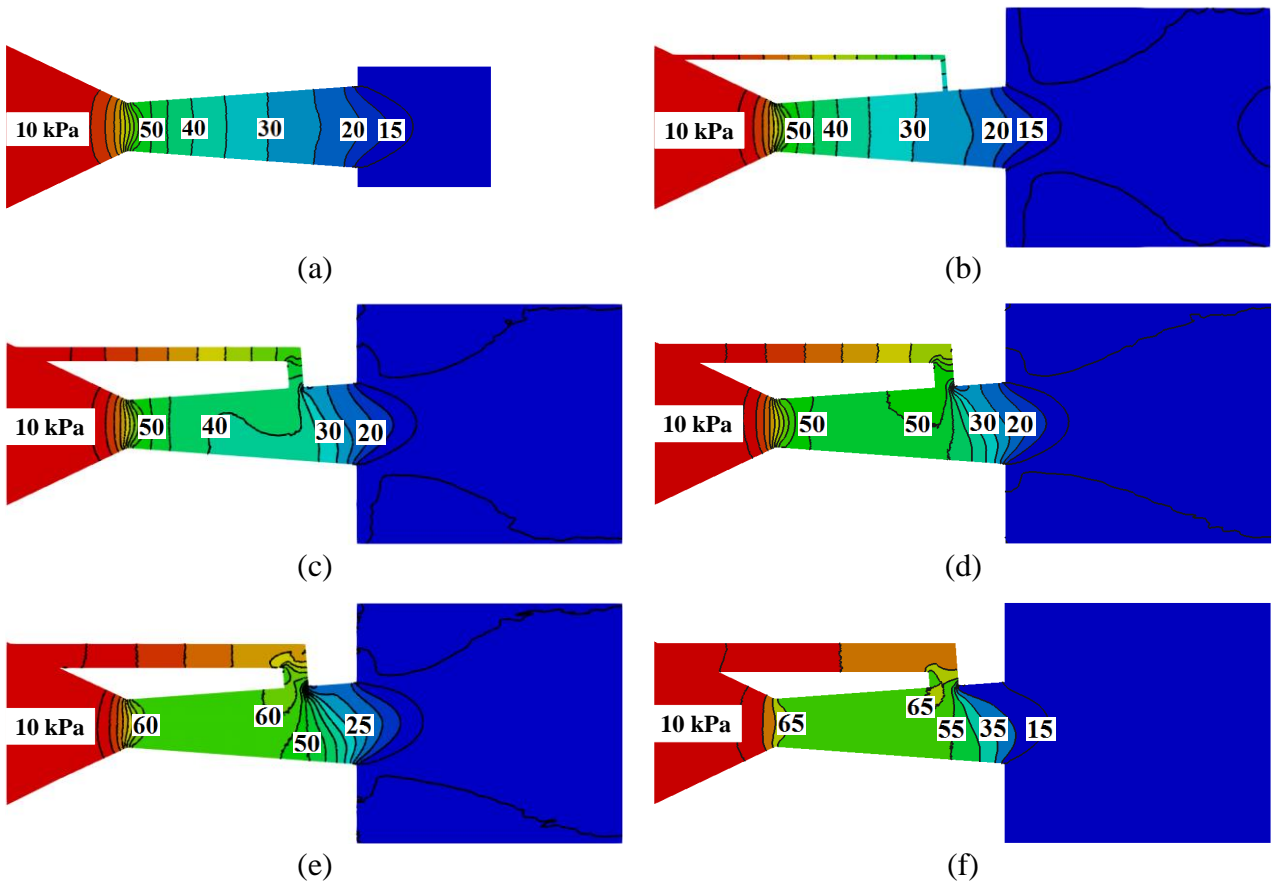

**Figure 12.** Pressure (kPa) contours for (a) baseline nozzle, (b) $h/H_t = 0.1$, (c) $h/H_t = 0.3$, (d) $h/H_t = 0.4$, (e) $h/H_t = 0.5$ and (f) $h/H_t = 0.6$ thrust vectoring nozzle at $P_{out} = 10$ kPa.

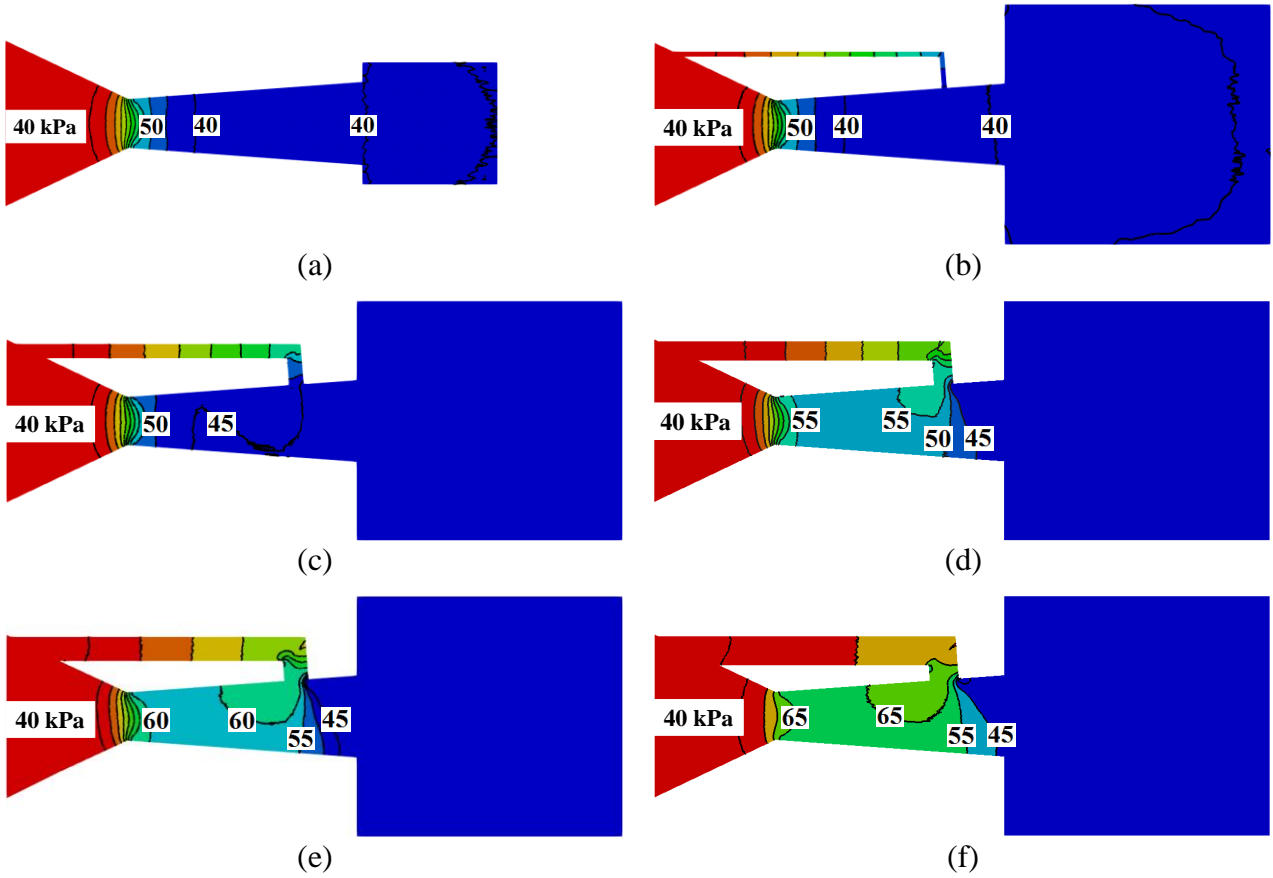

**Figure 13.** Pressure (kPa) contours for (a) baseline nozzle, (b) $h/H_t = 0.1$, (c) $h/H_t = 0.3$, (d) $h/H_t = 0.4$, (e) $h/H_t = 0.5$ and (f) $h/H_t = 0.6$ thrust vectoring nozzle at $P_{out} = 40$ kPa.

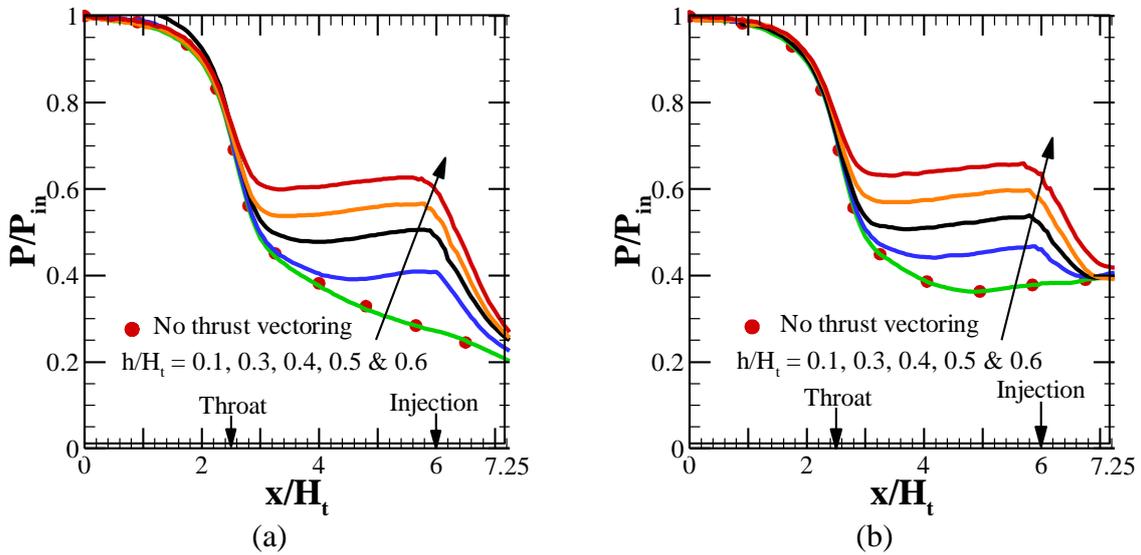

**Figure 14.** Variation of the normalized pressure along the centerline at (a) $P_{out} = 10$ kPa and (b) $P_{out} = 40$ kPa for different bypass channel dimensions.

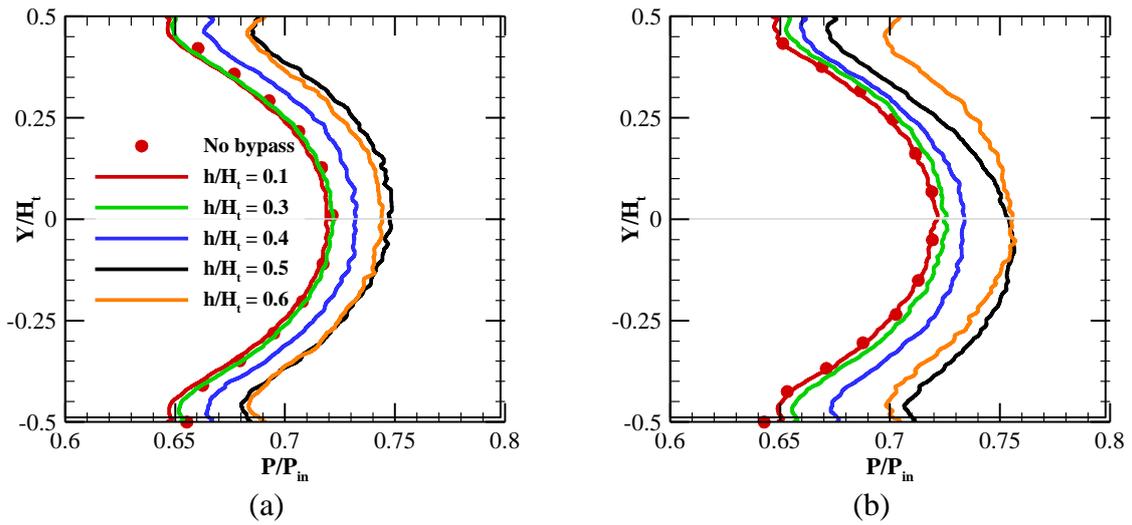

**Figure 15.** Variation of the normalized pressure along the throat at (a) $P_{out}$ = 10 kPa and (b) $P_{out}$ = 40 kPa for different bypass channel dimensions.

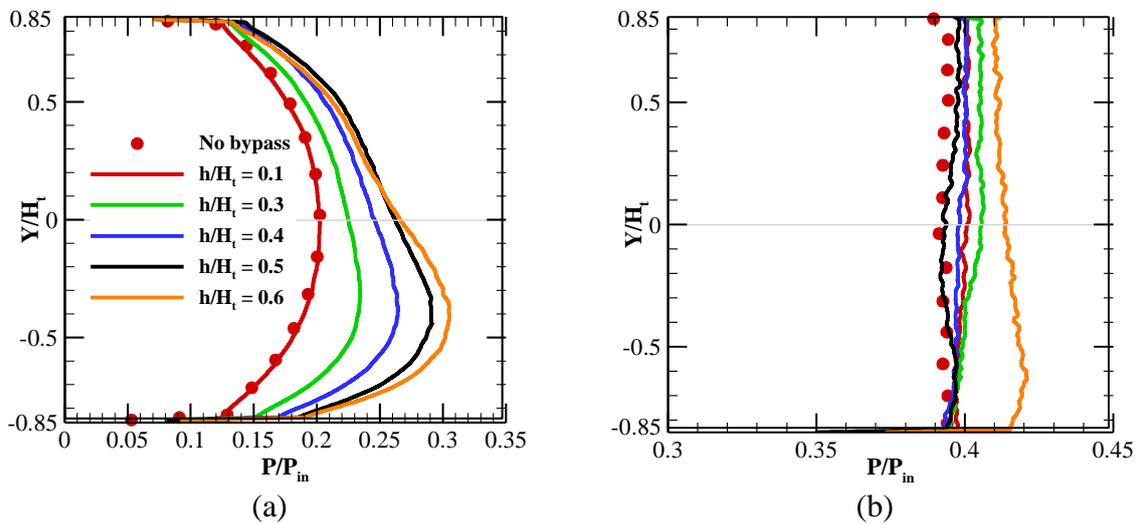

**Figure 16.** Variation of the normalized pressure along the exit plane at (a) $P_{out}$ = 10 kPa and (b) $P_{out}$ = 40 kPa for different bypass channel dimensions.

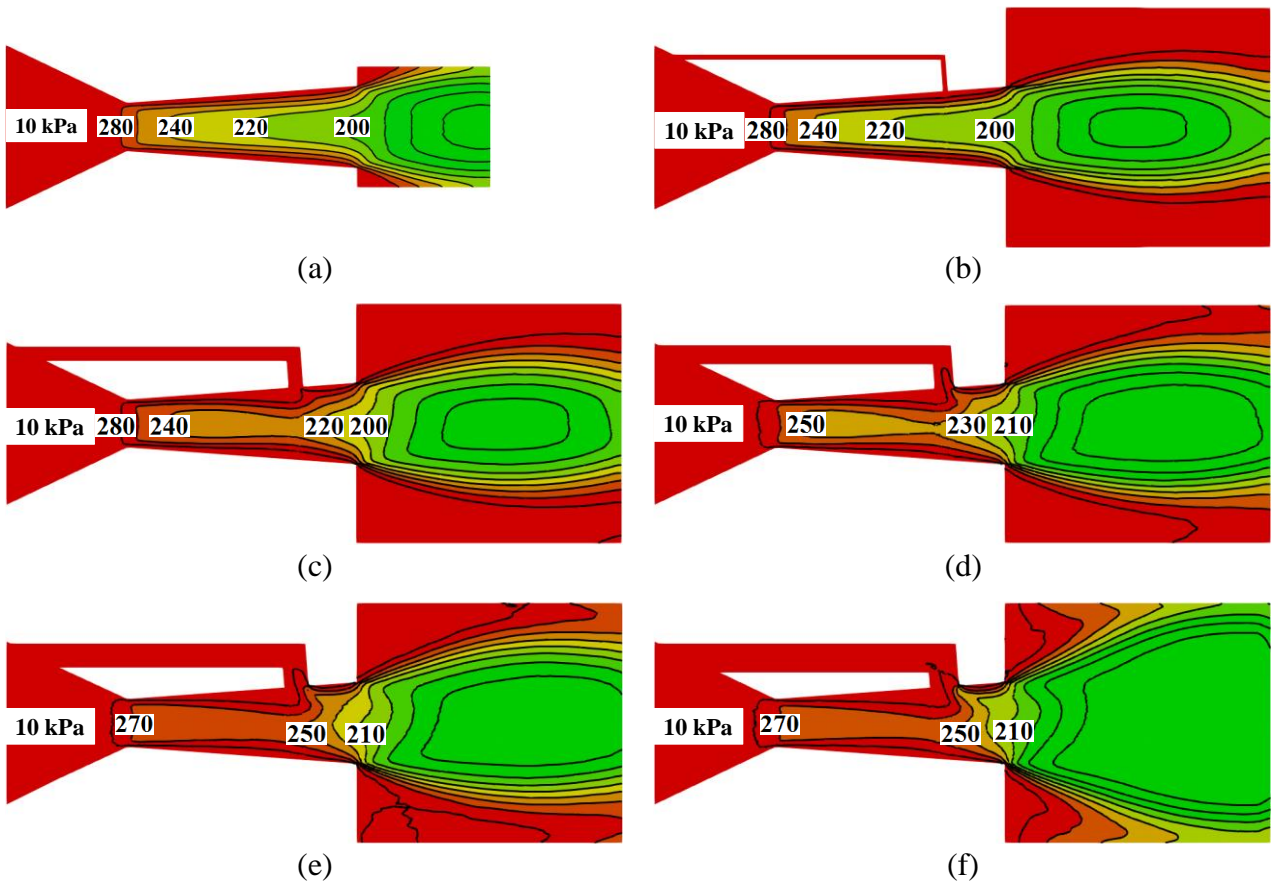

**Figure 17.** Temperature (K) contours for (a) baseline nozzle, (b) $h/H_t = 0.1$, (c) $h/H_t = 0.3$, (d) $h/H_t = 0.4$, (e) $h/H_t = 0.5$ and (f) $h/H_t = 0.6$ thrust vectoring nozzle at $P_{out} = 10$ kPa.

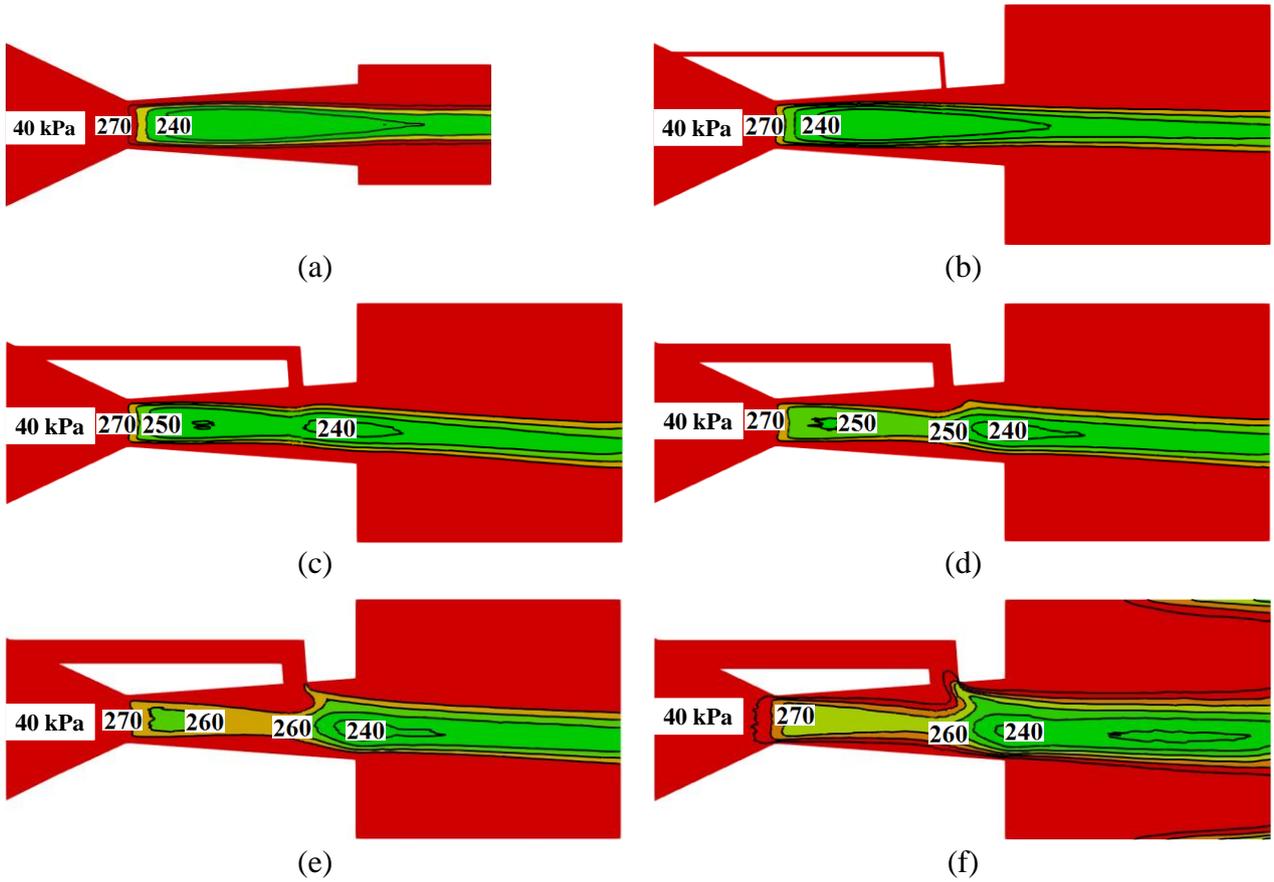

**Figure 18.** Temperature (K) contours for (a) baseline nozzle, (b) $h/H_t = 0.1$, (c) $h/H_t = 0.3$, (d) $h/H_t = 0.4$, (e) $h/H_t = 0.5$ and (f) $h/H_t = 0.6$ thrust vectoring nozzle at $P_{out} = 40$ kPa.

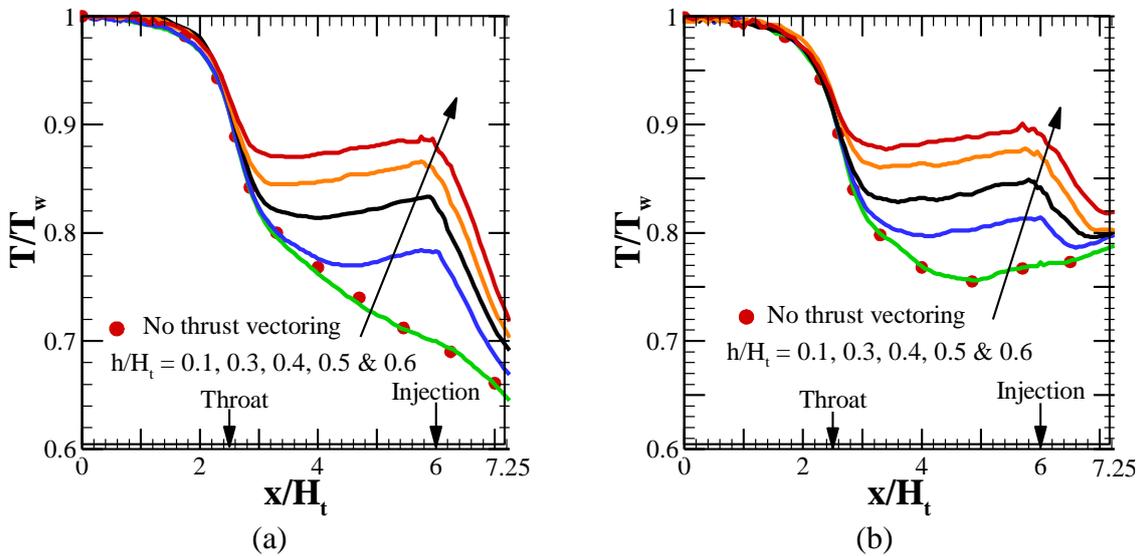

**Figure 19.** Variation of the normalized temperature along the centerline at (a) $P_{out} = 10$ kPa and (b) $P_{out} = 40$ kPa for different bypass channel dimensions.

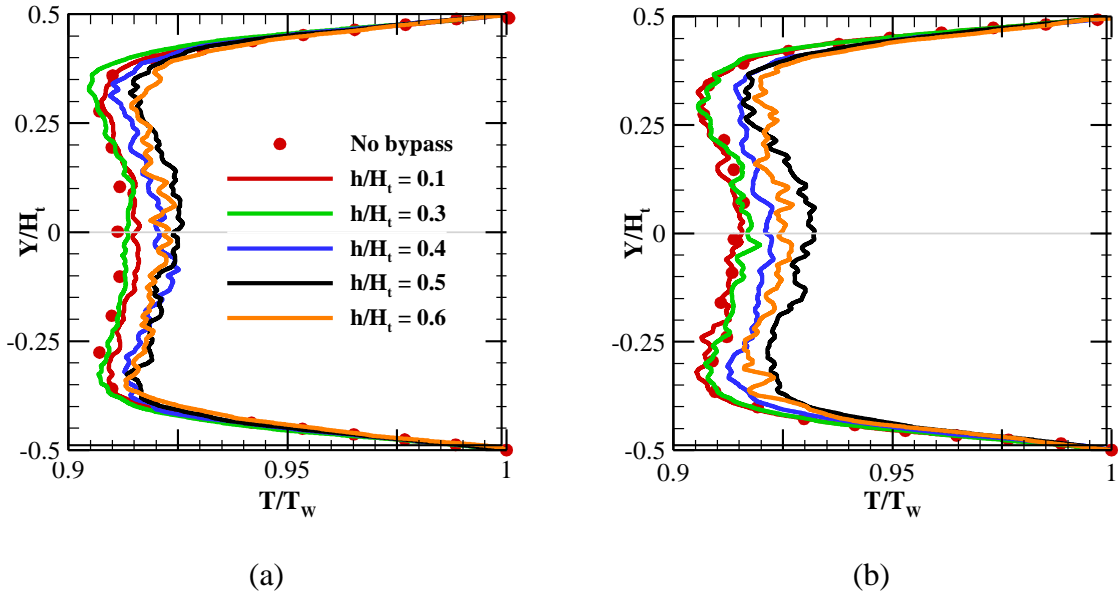

**Figure 20.** Variation of the normalized temperature along the throat at (a) $P_{out}$ = 10 kPa and (b) $P_{out}$ = 40 kPa for different bypass channel dimensions.

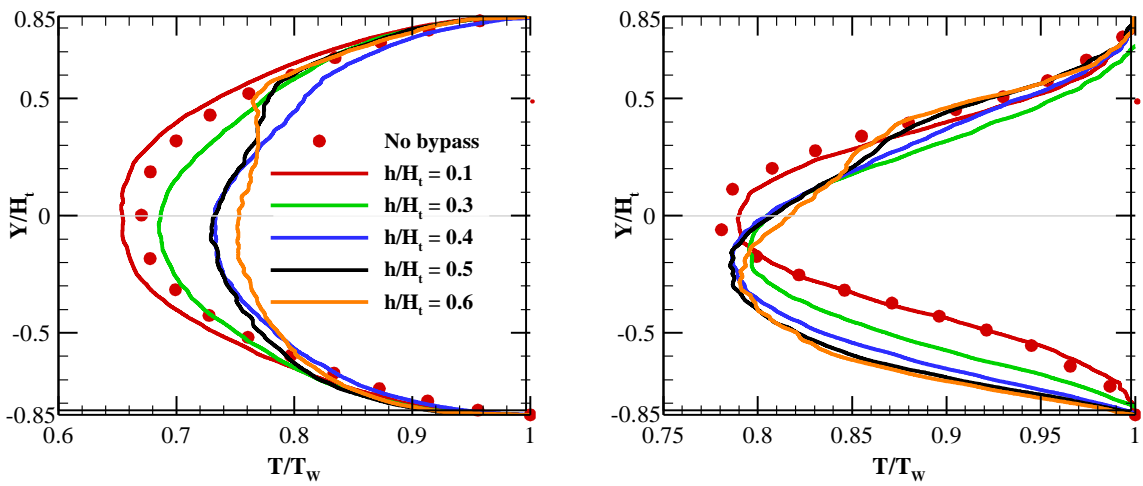

**Figure 21.** Variation of the normalized temperature along the exit at (a) $P_{out}$ = 10 kPa and (b) $P_{out}$ = 40 kPa for different bypass channel dimensions.

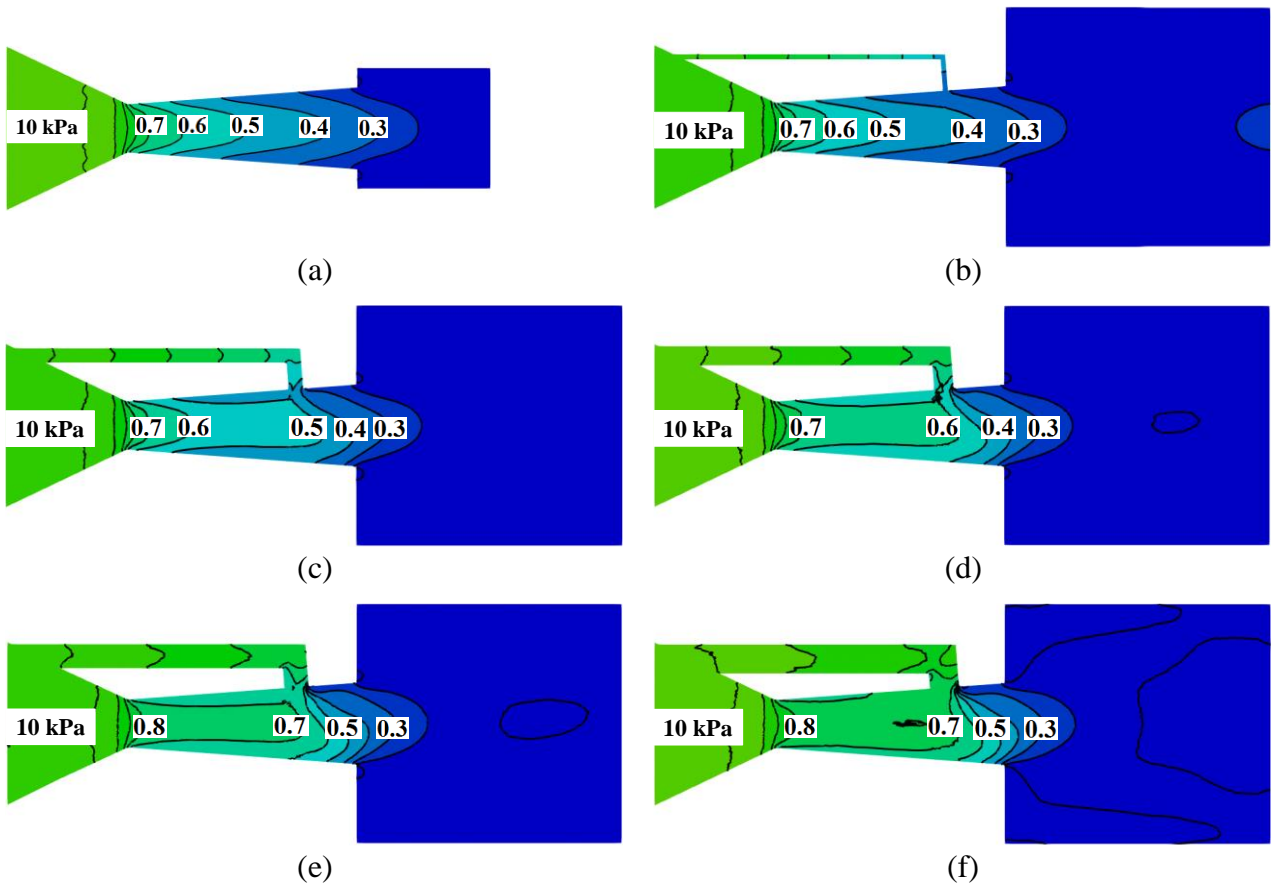

**Figure 22.** Density (kg/m³) contours for (a) baseline nozzle, (b) $h/H_t = 0.1$, (c) $h/H_t = 0.3$, (d) $h/H_t = 0.4$, (e) $h/H_t = 0.5$ and (f) $h/H_t = 0.6$ thrust vectoring nozzle at $P_{out} = 10$ kPa.

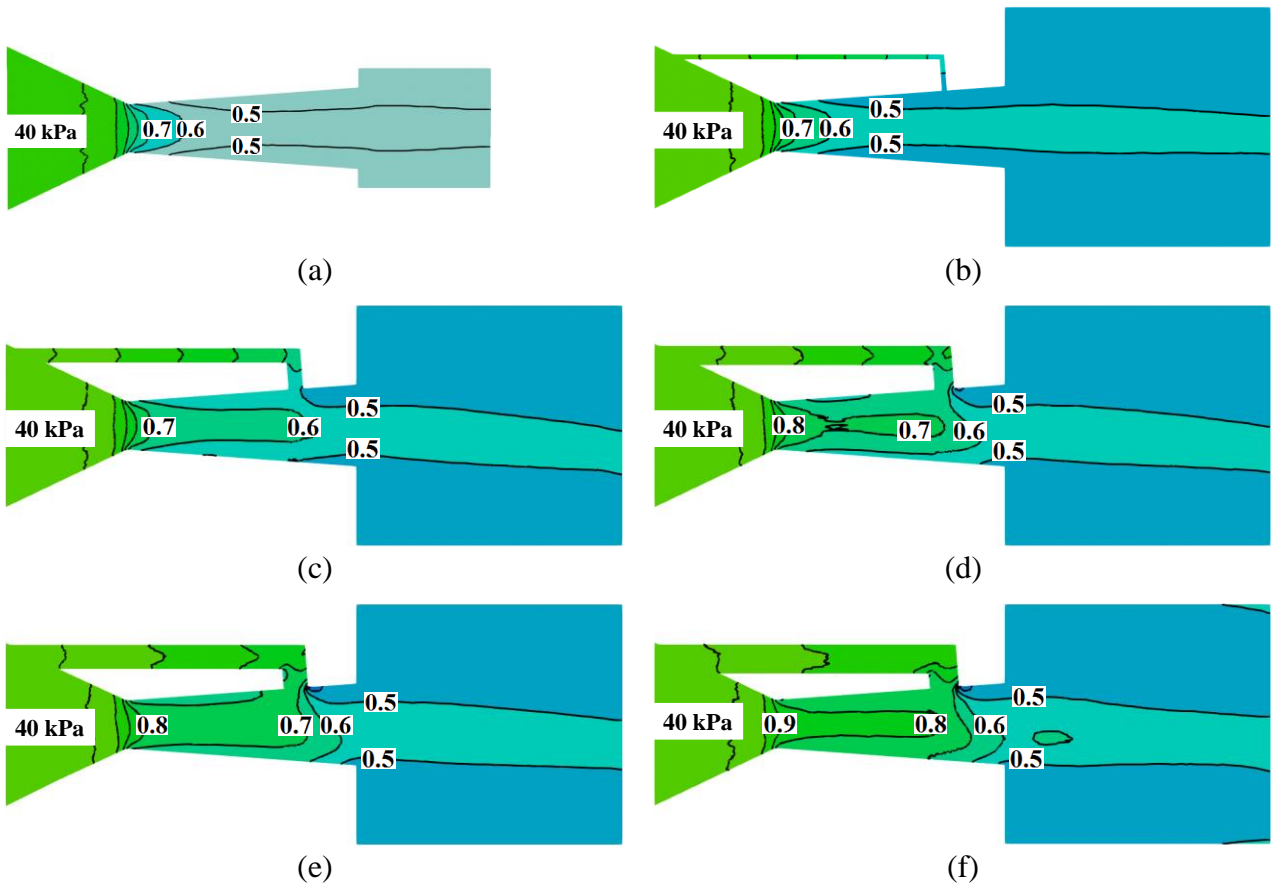

**Figure 23.** Density (kg/m$^3$) contours for (a) baseline nozzle, (b) $h/H_t = 0.1$, (c) $h/H_t = 0.3$, (d) $h/H_t = 0.4$, (e) $h/H_t = 0.5$ and (f) $h/H_t = 0.6$ thrust vectoring nozzle at $P_{out} = 40$ kPa.

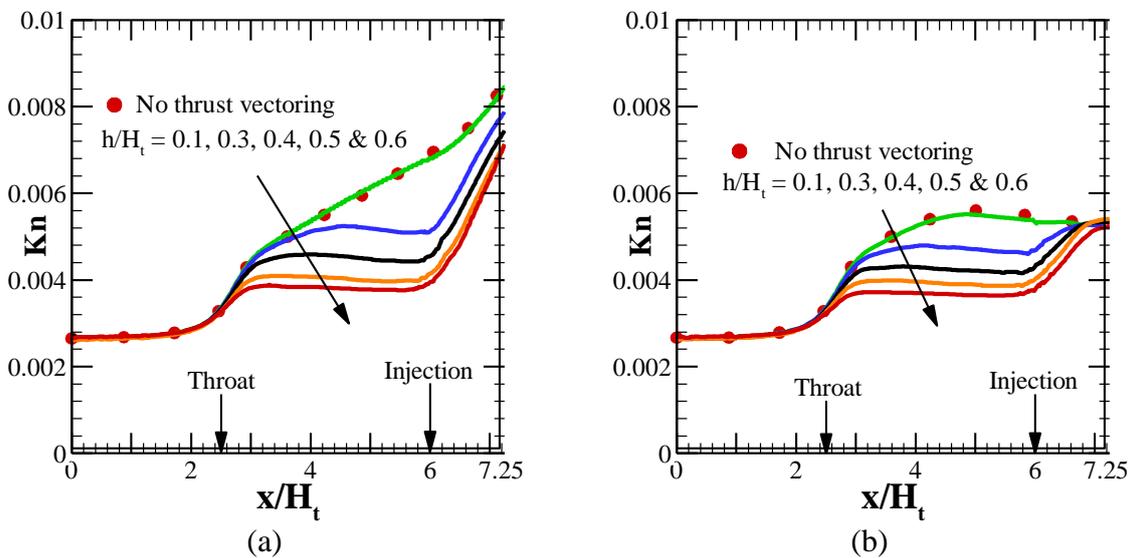

**Figure 24.** Variation of the Knudsen number along the centerline at (a) $P_{out} = 10$ kPa and (b) $P_{out} = 40$ kPa for different bypass channel dimensions.

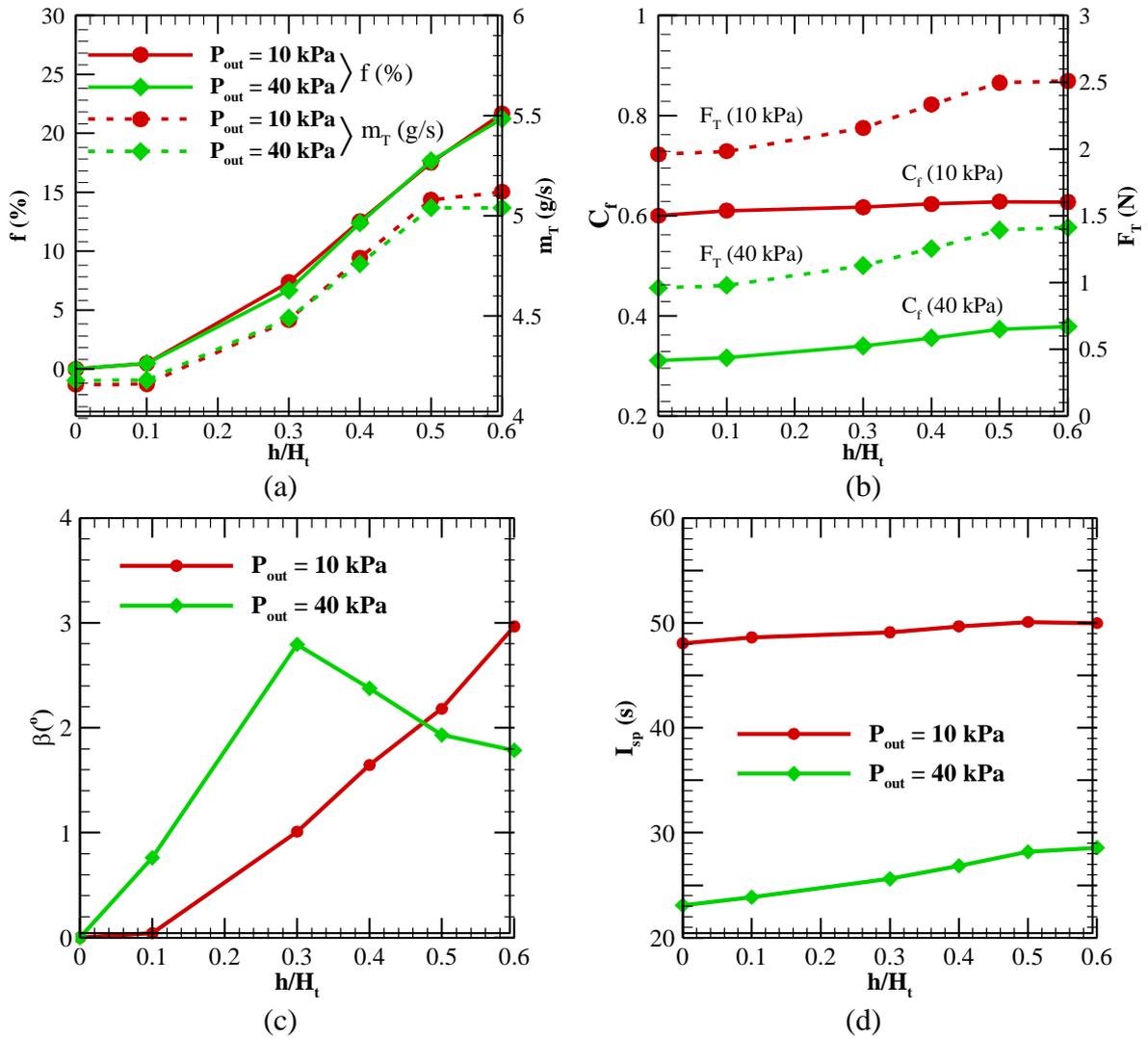

**Figure 25.** Variation of (a) total mass flow rate and secondary flow percentage, (b) thrust force and thrust coefficient, (c) thrust vectoring angle and (d) specific impulse for different bypass dimension at $P_{out}$ = 10 and 40 kPa.